\RequirePackage{lineno}
\documentclass[floatfix, onecolumn, prb]{revtex4}

\usepackage{hyperref}
\usepackage{color}
\usepackage{multirow}
\usepackage{epsfig}
\usepackage{graphicx}
\usepackage{amsmath}
\usepackage{amssymb}
\usepackage{bm}
\usepackage{dcolumn}
\usepackage{braket}
\usepackage{bbold}

\usepackage{ulem}
\usepackage{natbib}
\usepackage{booktabs}
\usepackage{graphicx}

\usepackage{lineno}
\usepackage{pdfpages}
\begin{document}
	
\title{High-throughput design of all-$d$-metal Heusler alloys for magnetocaloric applications}
	
\author{Nuno~M.~Fortunato$^1$}
\email{fortunato@tmm.tu-darmstadt.de}
\author{Xiaoqing Li$^2$}
\author{Stephan Sch\"onecker$^2$}
\author{Ruiwen~Xie$^1$}
\author{Andreas Taubel$^1$}
\author{Franziska Scheibel$^1$}
\author{Ingo~Opahle$^1$}
\author{Oliver~Gutfleisch$^1$}
\author{Hongbin~Zhang*$^1$} 
\affiliation{$^1$Institute of Materials Science, TU Darmstadt, 64287 Darmstadt, Germany}
\affiliation{$^2$ Department of Materials Science and Engineering, KTH - Royal Institute of Technology, Stockholm, SE-10044, Sweden}

\maketitle

Due to their versatile composition and customizable properties, A$_2$BC Heusler alloys have found applications in magnetic refrigeration, magnetic shape memory effects, permanent magnets, and spintronic devices. The discovery of all-$d$-metal Heusler alloys with improved mechanical properties compared to those containing main group elements, presents an opportunity to engineer Heuslers alloys for energy-related applications. Using high-throughput density functional theory calculations, we screened magnetic all-$d$-metal Heusler compounds and identified 686 (meta)stable compounds. Our detailed analysis revealed that the inverse Heusler structure is preferred when the electronegativity difference between the A and B/C atoms is small, contrary to conventional Heusler alloys. Additionally, our calculations of Pugh ratios and Cauchy pressures demonstrated that ductile and metallic bonding are widespread in all-$d$-metal Heuslers, supporting their enhanced mechanical behaviour. We identified 49 compounds with a double-well energy surface based on Bain path calculations and magnetic ground states, indicating their potential as candidates for magnetocaloric and shape memory applications. Furthermore, by calculating the free energies, we propose that 11 compounds exhibit structural phase transitions, and propose isostructural substitution to enhance the magnetocaloric effect.
	
\section{Introduction}
	
\begin{figure*}
	\includegraphics[width=0.9\textwidth]{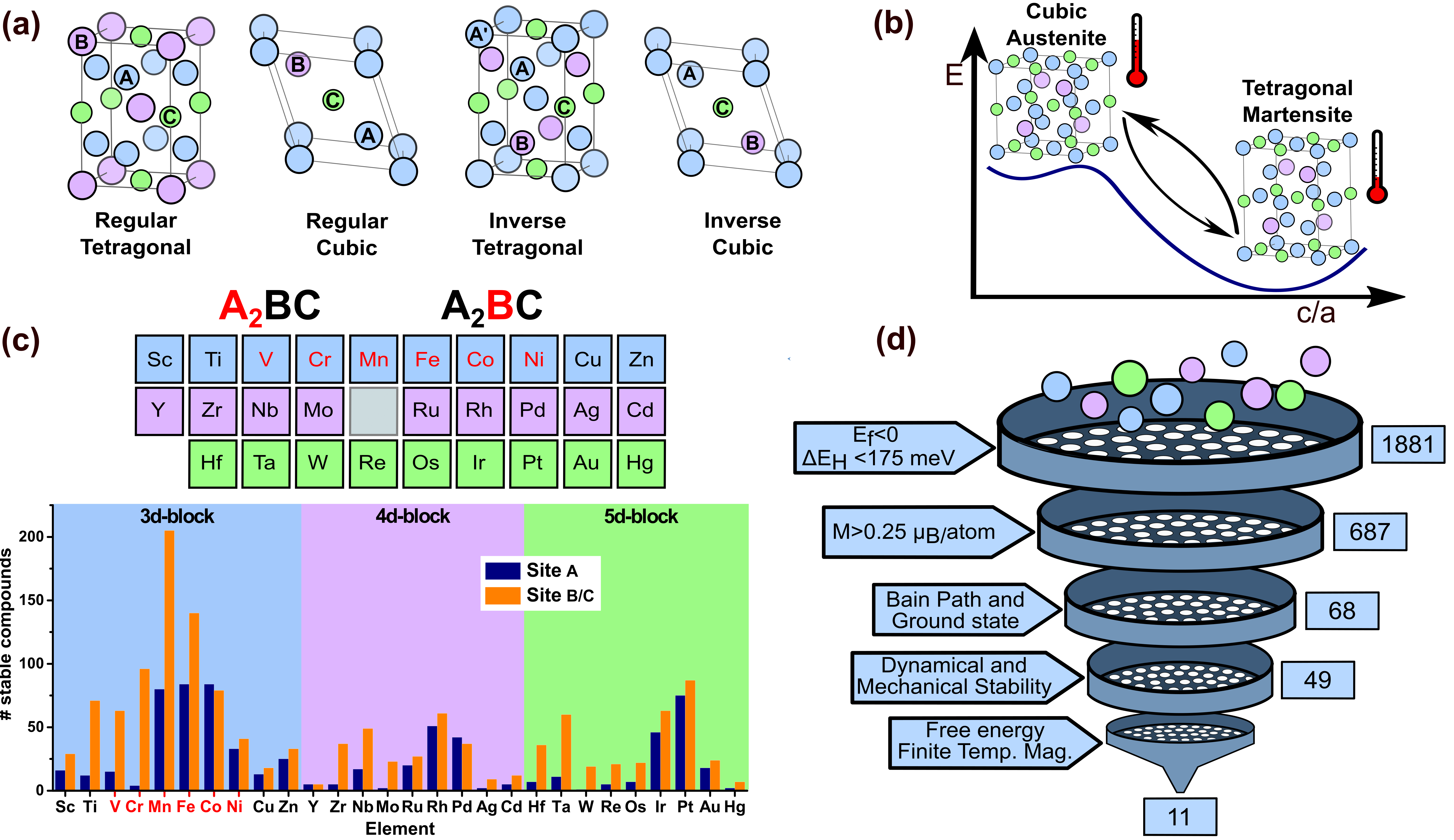}
	\caption{Figure 1 a: The crystal structures in our HTP search, the inverse/regular cubic Heuslers in their primitive setting and the inverse/regular in their tetragonal setting. b: A schematic of how the Bain path relates to the structural transition of low temperature tetragonal (ground state) martensite to the  high-temperature cubic austenite. c: The compositional phase space of our HTP, the graph below shows the frequency of elements in our stable magnetic compounds in the resulting screening. d: The summary of our HTP results showing each step of the screening procedure and the respective number of compounds remaining. 
	}
	\label{fig:periodic}
\end{figure*}

Heusler alloys exhibit a wide range of stable compounds with an A$_2$BC stoichiometry, forming a crystal structure comprised of rock salt (AB and AC) and cubic (AA' and BC) sublattices \cite{Graf2011}. 
This leads to a wide spectrum of physical properties and promising applications, {\it e.g.}, magnetic refrigeration \cite{Gutfleisch2011,Gutfleisch2016a}, permanent magnets \cite{Matsushita2017},  
magnetic shape memory (MSM) effect \cite{Planes2009}, and spintronic devices \cite{Palmstrom2016}.
Conventionally, A and B are metallic elements and C is a main group element, resulting in the coexistence of covalent, ionic and metallic bonds\cite{Graf2011}\cite{Wollmann2017}.
From the structural point of view, in addition to the regular Hesuler structure with  L2$_1$-type (Fm$\overline{3}$m, spg. nº 225), 
Heusler alloys also occur in the XA-type (F$\overline{4}$3m, spg. nº 216), dubbed as inverse Heuslers, along with the tetragonal variants of both structures,  {\it i.e.},  in the I4/mmm (spg. nº 139) and I$\overline{4}$m2 (spg. nº 119) space groups, respectively.
 The inverse Heusler type can be obtained by exchanging the occupation of one of the A sites with the B-site element, resulting in two inequivalent A sites, as exemplified in Figure \ref{fig:periodic} (a). 
 Burch {\it et al.} observed that in Fe$_{3-x}$TM$_x$Si (TM = transition metal) the TM dopant adopts the A Fe-sites if it belongs to the same or later periodic group than Fe (similar occupation to inverse Heusler), while earlier elements prefer the B Fe-sites (as in regular Heuslers alloys), this is known as Burch's rule \cite{T.J.Burch1974}.
Furthermore, the relative stability between tetragonal and cubic structures in Heusler alloys, as described by the Bain path \cite{Bain1924}, can lead to a martensitic phase transition \cite{Entel2006a}, as sketched in Figure \ref{fig:periodic} (b). 
For instance,  Ni-Mn-X (X = Al, Ga, In, Sn) Heusler alloys  host a first-order structural transition where the metastable cubic Fm$\overline{3}$m structure acts as a high-temperature (austenite) phase that on cooling transforms to the ground state I4/mmm martensite phase and also lower symmetry modulated martensite phase can occur dependent dependent on the chemical composition \cite{Entel2018a}. A magnetostructural transformation involves both a structural change between the martensite and austenite phases, and a change in magnetic state, for instance the transition from paramagnetic (PM) to ferromagnetic (FM) in Co-Ni-Mn-In \cite{Gottschall2016}. A magnetostructural transformation can be induced by an externally applied magnetic field, owing to the stabilization of the FM phase, which can be used for magnetocaloric cooling. A further interesting application are magnetic shape memory alloys (MSM).

 More recently, Wei {\it et al.} \cite{Wei2015a} synthesised Ni$_2$Mn$_x$Ti$_{1-x}$ Heusler alloys, which are notably composed only of $d$-metals,  omitting the main group elements on the C site. 
 As in Ni-Mn-X alloys, the magnetostructual transition in Ni$_2$Mn$_x$Ti$_{1-x}$ can be tailored by the Mn-content, which also changes the valence electron per atom ratio (e/a). In addition Co-doping can be used to stabilize the FM state and increase the magnetization \cite{Wei2015a,Taubel2020}. Of note, the all-$d$-metal Co-Ni-Mn-Ti alloy exhibits better mechanical stability during repeated structural transitions, in comparison to the more brittle conventional Ni-Mn-X Heuslers, which has been attributed to the $d$-$d$ hybridization \cite{Yan2019,Cong2019,Wei2016,NevesBez2019}. This makes all-$d$-metal alloys promising for MSMs, magnetocaloric, barocaloric, and elastocaloric applications.  

Thus all-$d$-metal Heuslers standout as a promising candidate system for cooling cycles based on the magnetocaloric effect (MCE), for example by exploiting hysteresis using a simultaneous application of stress and field \cite{Gottschall2018}. 
In addition, the tunability of the electronic structure of all-$d$-metal Heuslers also open up new functionalities and applications. For instance, Co$_2$VMn Heusler alloys exhibit a tunable eg orbital occupancy, which makes them interesting as electrocatalysts for oxygen evolution reactions \cite{Yu2021b}.
Thus, the all-$d$-metal Heuslers are a new twist in the Heusler family and have brought fresh attention to this material class, in terms of as potential novel materials and their technological applications \cite{DePaula2021}.  

High-throughput (HTP) screening based on density-functional theory (DFT) has been widely used to design Heusler alloys by exploring their versatile chemical space \cite{Jiang2021,Zhang2020}. 
For instance, Sanvito {\it et al.} \cite{Sanvito2017} performed a screening of compositions with $d$-block and main group elements, finding 248 compounds that were labelled as thermodynamically stable. From these several candidates were synthesized as a trial: Co$_2$MnTi, Mn$_2$PtPd, Mn$_2$PtCo, and Mn$_2$PtV. Of note, the tetragonally distorted structures were only considered in the 248 compounds that were stable in the cubic phase. Moreover, Balluff and colleagues \cite{Balluff2017} carried out screening of the anti-ferromagnetic (AFM) ground states of Heusler compounds for spintronic applications, and predicted 21 compounds with Néel temperatures (T$_{\textrm{N}}$) above room temperature.

Additionally, there have been extensive HTP screenings focusing on the non-magnetic Heusler alloys and their possible applications, such as Heuslers as precipitates/alloy systems to improve the mechanical behaviour \cite{Kirklin2016} and energy generation using the thermoelectric effect \cite{He2016}. 
While Faleev and colleagues \cite{Faleev2017} used HTP calculations to argue that tetragonality correlates with features in the electronic structure of the corresponding cubic structures.
A different approach was made by Oliynyk et. al \cite{Oliynyk2016} that constructed a machine-learning model trained on crystallographic databases to predict the synthesizability of regular and inverse Heusler alloys, where new Ru-Ga-X (X = Ti-Co) Heusler alloys were successfully synthesized based on their models.

We have recently published the anomalous Hall and Nernst conductivity of FM all-$d$-metal regular Heusler compounds \cite{alldHall}, and in this work we detail the HTP screening of magnetic all-$d$-metal Heuslers. We shed light on the compositional range, ground state structures, and in particular, the mechanical and thermodynamic behaviour of all-$d$-metal Heuslers, based on a workflow for the discovery of materials for energy applications, namely MCE and MSM alloys \cite{MMX2023}. From the thermodynamic stability, we predict 686 (meta-)stable compounds and identify their magnetic/structural ground states. Using such an extensive dataset, we rationalize how intermetallic bonding affects the structural preference between regular and inverse Heusler structures as dictated by Burch's rule, and correlate the bonding behaviour with the mechanical behaviour. Furthermore, following the concept of the Bain path and the relative stability of the tetragonal and cubic phases, we screen such compounds for possible magneto-structural transitions by evaluating the their finite temperature magnetic and structural behaviour. Finally, we provide a path for the tuning of phase transition behaviour necessary for optimal MCE applications.  
This paves the way to design and optimize novel functional materials for (magnetic) shape memory and (magneto-)caloric applications. 

	\section{Results}

  \subsection{Thermodynamic stabilities}  
   
The compositional range of our HTP screening covers the whole $d$-element block (excepting Tc) with at least one magnetic 3$d$ atom (V, Cr, Mn, Fe, Co, Ni) on either the A or B/C sites (see Figure \ref{fig:periodic} (c)), resulting in a total of 5802 unique compositions. For each composition we consider four different structures: the primitive four-atom regular and inverse Heusler cells ({\it i.e.}, L2$_1$ and XA types) and the two respective 8-atom tetragonally distorted structures ({\it i.e.} with space groups of I4/mmm and I$\overline{4}$m2), in both the FM and non-magnetic settings. In total, we find 2482 ternary compositions with a negative formation energy (E$_\textrm{f}$) and 1881 with a distance to the convex Hull ($\Delta$E$_\textrm{H}$) lower than 175 meV/atom. 
The frequency of occurrence for each element in those 1881 compounds is shown in Fig.\ref{fig:periodic} (c).
Obviously, among the 3$d$ magnetic elements, Fe, Co, and Mn are the most common elements due to the constraint of the compositional space. 
Whereas among the 4$d$ and 5$d$ blocks, there are two maxima that occur at groups 5 (Nb,Ta) and 9-10 (Rh,Ir, Pd, and Pt) tapering off at the late, early and in-between elements.  

	Strictly speaking, a system is expected to be stable if it is on the convex Hull ($\Delta$E$_\textrm{H}$ = 0), however, in practice a cut-off is usually allowed, so that the metastable phases are also considered \cite{Jiang2021}. To ensure interesting magnetic properties, we focus only on compounds where either the cubic or tetragonal settings are magnetic, chosen as a magnetic moment larger than $0.25$ $\mu_B$.atom$^{-1}$, further reducing to 686 stable ternary all-$d$-metal Heusler compounds. With more than half below a 75 meV/atom distance to convex Hull (cf. Table S.1).
 \par
Our results can be well validated with the experimentally available cases (see Table S.2 for a comparison). For instance, the L2$_1$ structure of Ni$_2$MnTi has a large $\Delta$E$_\textrm{H}$ of 117 meV/atom, which can be attributed to the fact that experimental phase is a B$_2$-disordered phase \cite{Koch22}\cite{Taubel2020} and finite temperature effects.
Likewise, Co$_2$VMn reported by Yu {\it et al.}. \cite{Yu2021b}  is 67 meV/atom above the convex hull, again indicating that the all-$d$-metal Heuslers can still be synthesized even if they have a relatively large $\Delta$E$_\textrm{H}$.
At the same time, both Mn$_2$PdPt and Co$_2$MnTi are on the convex hull, and we find good agreement with the available experimental lattice parameters (see Table S.2) \cite{Sanvito2017}.
While Mn$_2$PtCo exhibits an AFM ordering \cite{Sanvito2017} leading to a small deviation in the lattice parameter, that is improved by considering FM states, as done below. 
Of note, Mn$_2$PtCo and Mn$_2$PtV were found to decompose to binaries by Sanvito {\it et al.} \cite{Sanvito2017}, with the former being 10 meV/atom and the latter 28 meV/atom above the convex hull. 
Based on these cases and given the possibility of AFM ordering (discussed below) that could lower the total energies, or disorder that adds a configurational entropy term \cite{Graf2009}, we consider that the 175 meV/atom tolerance of distance to the convex hull as a necessity to allow the inclusion of metastable compounds of interest.
 \par

\subsection{Bain Paths and magnetic ground states}
The magnetic ground state is a fundamental aspect for technological applications, for instance, the MCE in some Ni-Mn-X alloys involves AFM states, with strong dependence on the composition \cite{Entel2018a}. Meanwhile, different magnetic ground states can also lead to changes on the energy surface as a function of the c/a ratio, {\it i.e.} the Bain path. 
 Correspondingly, we set out to not only verify the magnetic ground state for both the cubic and tetragonal phases but also check its changes with respect to the c/a ratio.
 Taking the 686 stable magnetic compounds, the Bain paths are calculated for both the inverse and regular Heusler structures in the FM state, concurrently, we evaluate the AFM Bain paths for compounds labelled as AFM based on HTP screening of possible prototypical AFM states (cf. Figure S.1), as detailed in the Methods section.
  The resulting magnetic ground states, symmetries and lattice parameters for all the (meta-)stable compounds can be found in Table S.1.
By comparing the AFM and FM Bain paths, we find 151 systems with an AFM ground state, 114 in the regular and 37 in the inverse Heusler settings. Of those confirmed to have an FM ground state, 354
 (145) were regular (inverse) Heuslers. On the other hand, 468 out of 686 compounds are more stable with the tetragonal regular/inverse structure, which is inline with other studies that look at conventional Heusler alloys \cite{Faleev2017}, confirming that both the cubic and tetragonal structures should be considered in equal footing during HTP screening. Some compounds also prefer a ferrimagnetic arrangement, in particular inverse Heuslers tend to have a ferrimagnetic ordering when the A element is magnetic, due to its occupation of two distinct sublattices \cite{Graf2011}.
     \par
	
\subsection{Burch's rule in all-$d$-metal Heusler}       

\begin{figure*}
	\includegraphics[width=1.0\textwidth]{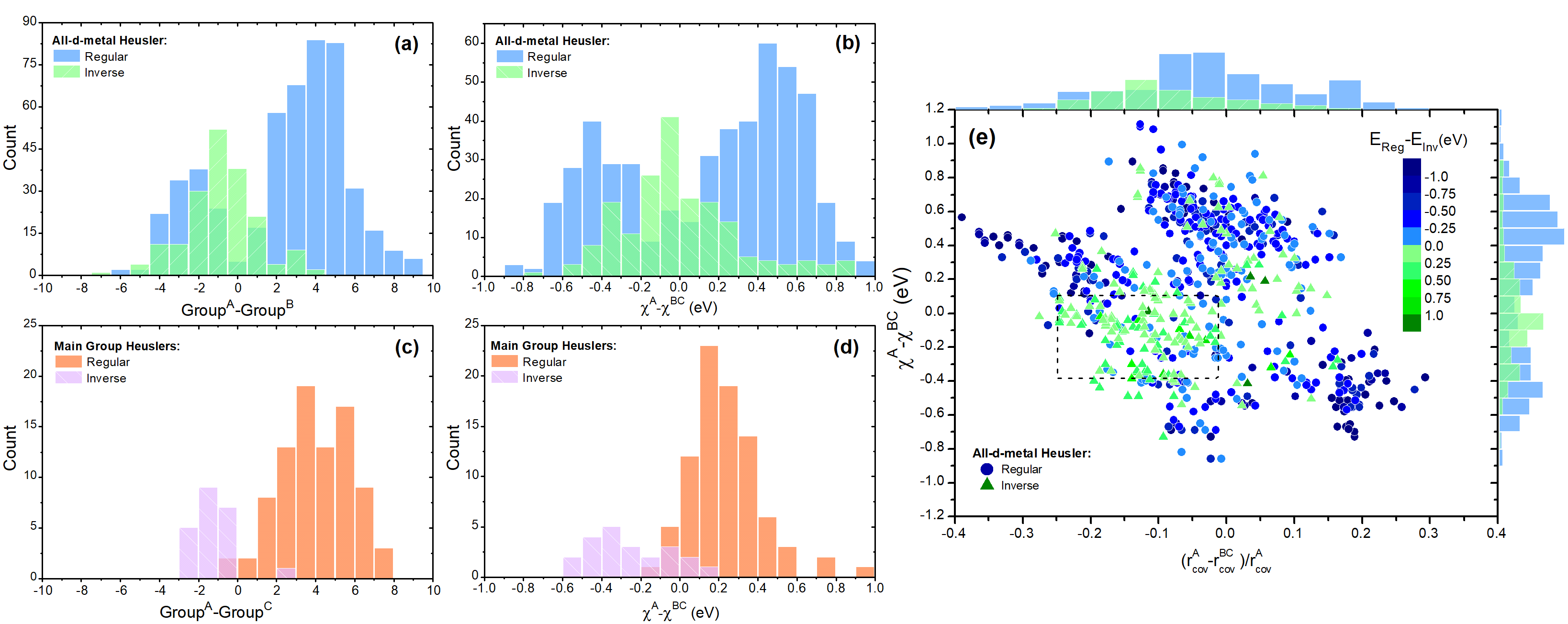}
	\caption{(a) Histogram showing the frequency of occurrence of regular (in blue) and inverse (green with stripes)  all-$d$-metal Heusler structures as function of the difference in group number of A and B elements. (b): The difference between the electronegativity of the A element and the average of B and C elements as a histogram. (c): The difference in group number of A and B elements applied to Heuslers with main group elements in literature. (d): The electronegativity difference histogram for Heusler with main group elements. (e): Map of the regular (circle) and inverse (triangle) structures as function of electronegativity and radius of A elements and the average of the B and C atoms. Colour indicates the energy difference between both structures, and dashed lines delineate the region where the inverse Heusler structure is most prevalent. On top of each axis we plot the respective histogram for the radius and electronegativity differences. }
	\label{fig:burch}
\end{figure*}

A fundamental question is whether a Heusler compound stabilizes in regular or inverse structures, and what is the underlying mechanism. According to Burch's rule, the inverse Heusler structure is preferred if the A element is an earlier transition metal (TM) than the B element for A$_2$BC Heuslers \cite{T.J.Burch1974}. The robustness of Burch's rule and DFT predictions was demonstrated by Kreiner {\it et al.}, who compared both to available experimental data on 90 phases \cite{Kreiner2014}. As such, our large HTP dataset presents an ideal opportunity to study structure preference in the all-$d$-metal Heusler family. To this end, we plot in Figure \ref{fig:burch}(a) the histogram of our stable regular/inverse Heusler structures with respect to the difference between the periodic group numbers of the A and B elements. Likewise, Figure S.2 (a) shows the distribution of the difference between the A and C element's group numbers, in the absence of main group elements we label C as the element with the highest group number (excluding the well defined A element). 
Clearly, there is a strong preference for the regular structure when the group number of the A element larger than that of other elements (large positive peak). However, the regular structure also dominates the extreme of the negative side (earlier period A element), with the inverse structure being preferred around the middle. In short, the inverse structure seems to prefer small differences between the group numbers of A and B/C elements, and the Burch's rule doesn't fully apply.
\par

To rationalize such results, we note that conventional Heusler compounds with a C main group element, Burch's rule implies that the regular structure is stabilized by the two rock-salt sublattices that coordinate the more electronegative/electropositive C/B elements with the A atom. In contrast, in the all-$d$-metal Heuslers the regular structure is predominant when there is a small difference between the group numbers of A and B/C, hence electronegativity difference tends to be smaller.
 As a way to describe the bonding in the inverse all-$d$-metal Heusler structures, we plot the electronegativity difference of A ($\chi$\textsuperscript{A}) and of the average of B and C ($\chi$\textsuperscript{BC}) in Figure \ref{fig:burch}(b). We observe that in all-$d$-metal Heuslers the inverse setting tends to be preferred when ($\chi$\textsuperscript{A})-($\chi$\textsuperscript{BC}) is between roughly 0.1 and -0.2 eV, with the regular Heusler exhibiting two peaks (around -0.45 and 0.45 eV). This can be understood by recalling the Van Arkel-Ketelaar triangle, which indicates that in metallic and covalent bonding small differences in the electronegativity values are expected, as opposed to the large difference characteristic of ionic bonding \cite{Norman}.
  Thus, in all-$d$-metal Heuslers the inverse structure is preferred under small electronegativity difference, which favours the metallic bonding of the A-B and A'-C sublattices. To compare, we plot 108 experimentally reported main group based Heusler alloys in Figure \ref{fig:burch}(c) (listed in Table S.3), it is clear that Burch's rule is verified based on the negative inverse Heusler peak.
   Contrasting with the previous all-$d$-metal inverse Hesulers, those with main group elements in Figure \ref{fig:burch}(b) are predominant when $\chi$\textsuperscript{A})-($\chi$\textsuperscript{BC}) is negative. Interestingly, there is a smaller peak in all-$d$-metal inverse Heuslers at -0.35 eV which matches the peak on the main group plot, hinting that there may be multiple overlapping bonding trends. We note that although we included tetragonal and cubic systems in the preceding analysis, the same trends are also verified in the subset of compounds that have a cubic ground state as plotted in Figure S.2 (b).

We further refine our explanation by noting that an inverse Heusler can be obtained by swapping the B element in the rock-salt with one of the simple-cubic A sites.
In this substitutional picture the covalent radius mismatch of the solvent and solute elements should be considered as a key criterion, according to the Hume-Rothery rules \cite{Rothery}.
As such, the difference between the covalent radius \cite{CRC} of the A element (r\rlap{\textsuperscript{A}}\textsubscript{cov}) and  of the average radii of B and C (r\rlap{\textsuperscript{BC}}\textsubscript{cov}) shouldn't be too large.
Putting the electronegativity and radius descriptors together we map the structural preference between inverse and regular all-$d$-metal Heuslers in Figure \ref{fig:burch}(e). It is clear that radius mismatch of the inverse structure is centred around 0.125, penalizing larger values. It is noted that outliers are expected, as we do not explicitly consider all bonds in the structures (such as AA' bonding), nor do we consider the valency of the elements. Nevertheless, we provided a simple picture as to the origin of inverse Heuslers in all-$d$-metal systems, while highlighting the metallic/covalent nature of the bonding in the all-$d$-metal Heusler family and explaining the violation of Burch's rules in all-$d$-metal Heusler alloys.

\subsection{Mechanical properties and stability}
To shed light on the mechanical stability and the unique mechanical properties of the all-$d$-metal Heusler alloys, we systematically evaluate their elastic constants (C$_{ij}$) and  derive their moduli from the Voigt-Reuss-Hill method \cite{Hill}.
From the ab-initio point of view, a key parameter is the Pugh ratio (B/G)\cite{Pugh}, which describes the expected degree of brittleness/ductility of a compound, where B (G) is the bulk (shear) modulus. Typically a reference value of 1.75 is taken as the start of ductile behaviour \cite{Niu2012}, however Wu  {\it et al.} \cite{Wu2019} remarked that Co-based cubic Heuslers are above this value but are still brittle, as such we discuss the Pugh ratio in relative terms instead of a fixed threshold.

 Another useful parameter, is Pettifor's Cauchy pressure, namely (C$_{13}$-C$_{44}$)/E and (C$_{12}$-C$_{66}$)/E for tetragonal systems, and C$_{12}$-C$_{44}$)/E for cubic systems \cite{DyCrys1998} \cite{Pettifor1992}, where E is the Young modulus. Here positive values signal a metallic (isotropic) character, whereas a negative value implies a more covalent/ionic (angular) bonding. Both Yan {\it et al.} \cite{Yan2019} and Cong \cite{Cong2019} {\it et al.} put forward that the higher Pugh ratio and larger positive Cauchy pressure of Ni-Mn-Ti in comparison to the conventional Ni-Mn-X Heusler alloys is a possible explanation for its better mechanical performance.

\begin{figure*}
	\includegraphics[width=0.9\textwidth]{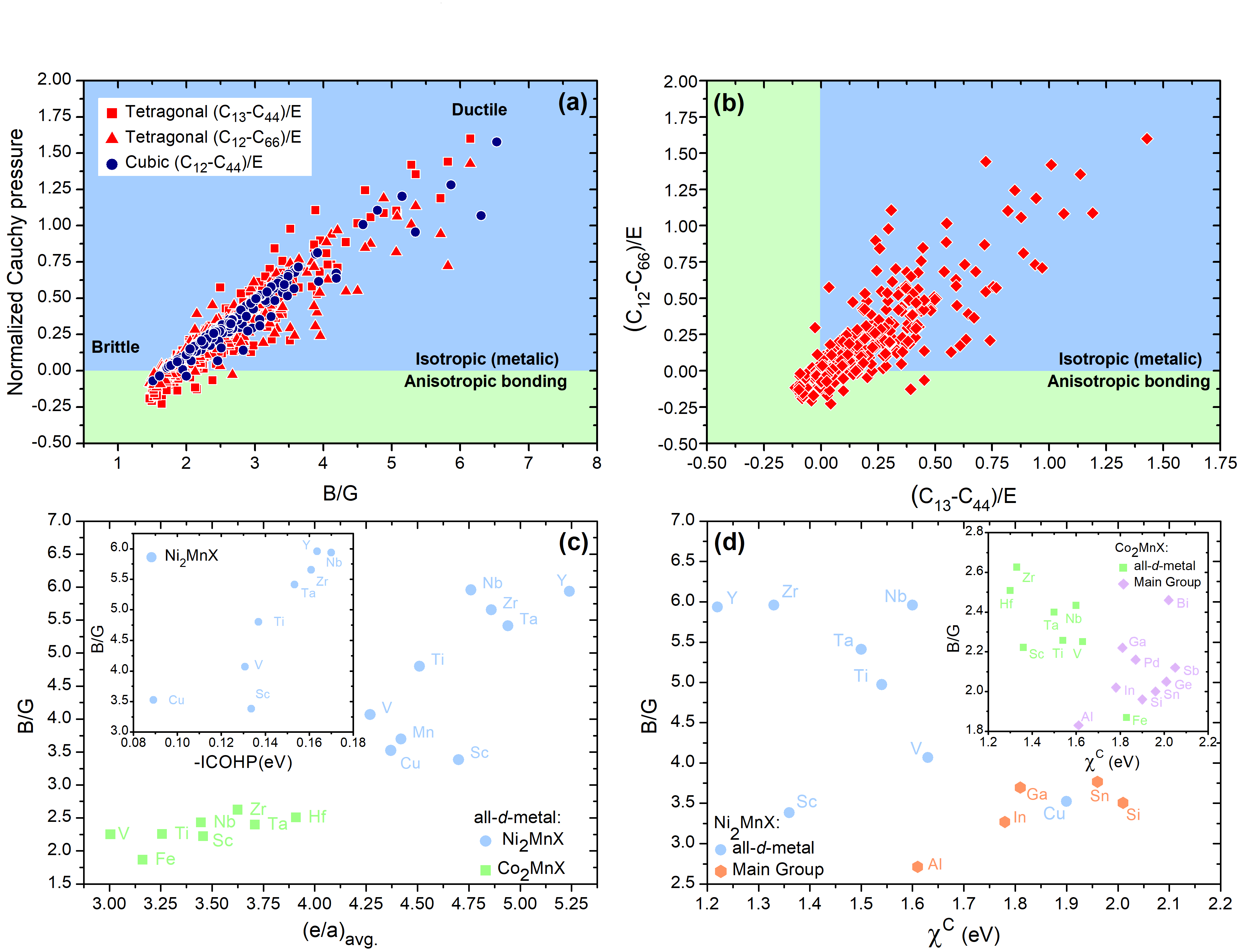}
	\caption{(a): The Pugh ratio showing the brittle/ductile behaviour versus the normalized Cauchy pressure which accounts for metallic or anisotropic nature. The cubic system is in blue circles and the two Cauchy pressures for the tetragonal  systems are plotted separately as red triangles and squares. (b): The two Cauchy pressures for tetragonal systems plotted against each other, clarifying how many tetragonal systems are isotropic in each one. (c) Pugh ratio of hypothetical and stable all-$d$-metal Ni$_2$MnX and Co$_2$MnX cubic Heuslers versus the compositional average of e/a, the inset graph shows how the Pugh ratio scales with the bonding strength measured by the total -ICOHP averaged up to the third nearest-neighbours. (d) The Pugh ratios of the  Ni$_2$MnX and Co$_2$MnX (in the inset) families compared to main group based compounds as a function of the electronegativity of the X element. Data for Co$_2$MnX main group systems taken from  \cite{Wu2019}. }
	\label{fig:mech}
\end{figure*}

To highlight the broader behaviour of the all-$d$-metal Heusler family we plot both parameters in Figure \ref{fig:mech} (a) and (b). It is immediately apparent that most compounds have large Pugh ratio, with a few being comparable with Ni$_2$MnTi (B/G=4.97). While, there are some (mostly tetragonal) that are brittle or with anisotropic bonding, most compounds are still labelled as metallic and ductile. These results suggest that the ductile/isotropic bonding behaviour is a widespread feature of the all-$d$-metal Heusler family, likely owing to the composition including only TM elements.
 
	Turning back to the mechanical behaviour of Ni$_2$MnTi, it is worth noting that  Ni$_2$MnX (X = Sn, Al, Ga, In, Si) alloys with main group elements already have large Pugh ratios (between 2.71 and 3.77). This is higher than many of the compounds in our dataset, thus high Pugh ratios are hallmark of the Ni$_2$MnX system, that is then enhanced by specific TM elements, such as Ti. For comparison, Co$_2$MnX Heuslers with main group elements, which have been explored for spintronics applications, showcase Pugh ratios ranging from 1.96 to 2.46  \cite{Wu2019}, with their all-$d$-metal counterparts not showing enhanced ductile behaviour, with ratios between 2.22 and 2.63. Heusler alloys are often discussed based on the number of valence electrons, for instance, the 24-electron-rule for semiconducting behaviour \cite{Graf2011} and e/a ratio when discussing structural transitions \cite{Taubel2020}. Given the metallic nature of all-$d$ Heuslers, we discuss their ductility in terms of the elemental e/a ratios provided by Mizutani {\it et al.} \cite{Mizutani2017}, which assigns a number to the electrons with an itinerant character from ab-initio methods. Figure \ref{fig:mech} (c) plots the compositional average (labelled as (e/a)$_{avg}$) of Ni$_2$MnX and Co$_2$MnX cubic Heuslers against their Pugh ratios. It is observed that ductile behaviour generally increases with the (e/a)$^{it.}$$_{avg}$ value, despite some deviation that is expected from the effect of local environment on the e/a. While conventional Ni$_2$MnX systems, seem insensitive to the e/a ratio (their e/a matches their valency), instead their ductility scales with increasing electronegativity as seen in Figure \ref{fig:mech} (d). While conversely in the all-$d$-metal compounds, the ductility increases with smaller electronegativity, attesting to a difference in the underlying bonding behaviour. This behaviour can be quantified using by integrating the Crystal Orbital Hamiltonian Populations (ICOHP) \cite{Dronskowski1993,Maintz2016,Maintz2013}, which partitions the DFT band-structure into bonding (negative COHP) or anti-bonding (positive COHP) contributions, based on the Hamiltonian matrix element of a given pair of orbitals. Interestingly, in all-$d$-metal Ni$_2$MnX the Pugh ratio increases with more negative ICOHP indicating stronger bonding (inset of \ref{fig:mech} (c)), with Ni-X and Mn-X contributions being the strongest contributions (not shown). All together this indicates that the stronger \textit{p-d} hybridization from a more electronegative main group element systems results in stronger covalent bonding, leading to more brittle behaviour. Conversely, stronger \textit{d-d} bonds in all-$d$-metal systems, due to the increasing number of itinerant electrons, result in a more ductile behaviour.

\par

In addition to the thermodynamic stability criteria (E$_f < 0$, and $\Delta$E$_\textrm{H}$ $<$ cut-off ), both the dynamical (from phonon band structure) and mechanical (based on elastic constants \cite{Mouhat2014}) stabilities should be evaluated, as described in the Methods section. 
Out of 686 compounds, most are stable in both criteria, with dynamical stability being the most stringent with 70 unstable compounds, and 33 mechanically unstable (c.f. Table S.4 and S.5).

Turning now to identifying Heusler compounds with the magneto-structural coupling needed for MCE and MSM, a typical double-well energy surface is expected along the Bain path, connecting a stable tetragonal and a (metastable) cubic structure \cite{Entel2006a}. 
Based on this criterion, we find 53 regular and 14 inverse Heusler compounds with a possible transition (aside from Ni$_2$MnTi). Among such 68 compounds, the most common case is FM states for both the martensite and austenite (c.f. Table S.6).  There are also 19 regular and 5 inverse systems exhibiting AFM ground state in either phase, with those compounds being mostly AFM in the martensite with a FM austenite, reflecting the previously observed trend. Regarding stability, the tetragonal martensite phase can be straightforwardly determined, {\it i.e.}, dynamical and mechanical instabilities belie the existence of other more stable structures. 
In contrast, for the cubic austenite structures, imaginary phonon modes are known to occur in experimentally realised compounds, {\it e.g.}, Ni$_2$MnGa, indicating that the martensite phase is more stable than the austenite phase \cite{Entel2006a}. 
Since the phonon spectra are obtained from T = 0 DFT calculations, thus we do not exclude the austenite phases with imaginary phonon modes, and discard only those compounds with unstable tetragonal structures. Based on these considerations, we find that 53 martensite structures are dynamically stable, while 62 mechanically stable, resulting in 49 fully stable martensite structures shown in Table \ref{Tab:TRAN}, their Bain paths can be seen in Figures S.3 and S.4.
Interestingly, several compounds are similar to systems with main group elements that have transitions, namely the Ni$_2$MnX, Mn$_2$NiX and Rh$_2$XY families \cite{Wu2018}. The energy difference between the austenite and martensite ranges from a few meV, up to several hundred meV.  Of note, Co$_2$VMn has been synthesized in the cubic phase \cite{Yu2021b}, despite the tetragonal state being 4 meV/atom lower than the cubic, indicating that small energy differences make determining a ground state difficult.

\begin{table*}[]\centering
	\caption{Thermodynamic stability of compounds with possible transition with a mechanically and dynamically stable martensite, along with the magnetic states and moments for both structures from our screening and the energy difference of the austenite and martensite. }
	\vspace{0.1cm}
	\begin{tabular*}{\linewidth}{@{\extracolsep{\fill}}cccccccccc@{}}
		\toprule
Composition & E$_\textrm{f}$ & $\Delta$E$_\textrm{H}$ & Mom. Tetra. & c/a & State  & Mom. Cub. & State  & E$_0^{\textrm{Cub.}}$-E$_0^{\textrm{Tetra.}}$ & Type \\
  & [eV/atom] & [eV/atom]  & [$\mu_B$/atom]  &   & Tetra.  & [$\mu_B$/atom] &  Cub. & [eV/atom] &  \\
				 \midrule
Au$_2$MnCu & 	 -0.059 &	 0.029 &	 1.03 &	 1.22 &	 FM &	 1.02 &	 FM &	 0.001 &	 Reg. \\ 
Fe$_2$PtZn & 	 -0.164 &	 0.114 &	 1.35 &	 1.33 &	 FM &	 1.34 &	 FM &	 0.003 &	 Inv. \\
Zn$_2$CrRh & 	 -0.117 &	 0.131 &	 0.92 &	 1.37 &	 FM &	 0.00 &	 AFM &	 0.004 &	 Reg. \\
Co$_2$VMn & 	 -0.099 &	 0.067 &	 0.23 &	 1.42 &	 FM &	 1.44 &	 FM &	 0.004 &	 Reg. \\
Zn$_2$MnPd & 	 -0.246 &	 0.070 &	 0.88 &	 1.27 &	 FM &	 0.89 &	 FM &	 0.007 &	 Reg. \\
Ir$_2$MnZn & 	 -0.195 &	 0.000 &	 0.00 &	 1.28 &	 AFM &	 0.78 &	 FM &	 0.008 &	 Reg. \\
Mn$_2$NbIr & 	 -0.170 &	 0.100 &	 0.00 &	 1.45 &	 AFM &	 1.18 &	 FM &	 0.008 &	 Reg. \\
Fe$_2$MnPt & 	 -0.078 &	 0.092 &	 0.00 &	 1.37 &	 AFM &	 2.25 &	 FM &	 0.009 &	 Inv. \\
Pd$_2$MnAg & 	 -0.146 &	 0.034 &	 1.11 &	 1.26 &	 FM &	 1.15 &	 FM &	 0.010 &	 Reg. \\
Mn$_2$MoPt & 	 -0.625 &	 0.000 &	 0.00 &	 1.34 &	 AFM &	 1.50 &	 FM &	 0.011 &	 Reg. \\
Zn$_2$CrPt & 	 -0.244 &	 0.094 &	 0.94 &	 1.31 &	 FM &	 0.83 &	 FM &	 0.011 &	 Reg. \\
Fe$_2$IrCo & 	 -0.019 &	 0.039 &	 0.00 &	 1.35 &	 AFM &	 2.15 &	 FM &	 0.013 &	 Reg. \\
Zn$_2$HfMn & 	 -0.082 &	 0.121 &	 0.36 &	 1.45 &	 FM &	 0.73 &	 FM &	 0.014 &	 Reg. \\
Fe$_2$VRh & 	 -0.128 &	 0.094 &	 0.95 &	 1.44 &	 FM &	 1.1 &	 FM &	 0.016 &	 Inv. \\
Rh$_2$MnCu & 	 -0.061 &	 0.021 &	 0.00 &	 1.32 &	 AFM &	 0.00 &	 AFM &	 0.016 &	 Reg. \\
Fe$_2$TaIr & 	 -0.239 &	 0.099 &	 1.01 &	 1.46 &	 FM &	 1.27 &	 FM &	 0.017 &	 Inv. \\
Cu$_2$ZrMn & 	 -0.014 &	 0.147 &	 0.53 &	 1.49 &	 FM &	 1.05 &	 FM &	 0.019 &	 Reg. \\
Ni$_2$MnCu & 	 -0.041 &	 0.030 &	 1.02 &	 1.34 &	 FM &	 1.06 &	 FM &	 0.020 &	 Reg. \\
Zn$_2$MnRh & 	 -0.249 &	 0.000 &	 1.02 &	 1.38 &	 FM &	 0.84 &	 FM &	 0.021 &	 Reg. \\
Ti$_2$MnAu & 	 -0.209 &	 0.151 &	 0.26 &	 1.50 &	 FM &	 0.45 &	 FM &	 0.027 &	 Inv. \\
Pt$_2$MnCu & 	 -0.309 &	 0.000 &	 1.01 &	 1.31 &	 FM &	 1.07 &	 FM &	 0.037 &	 Reg. \\
Ir$_2$MnCu & 	 -0.047 &	 0.065 &	 0.55 &	 1.32 &	 FM &	 0.57 &	 FM &	 0.038 &	 Reg. \\
Fe$_2$TiPt & 	 -0.399 &	 0.065 &	 1.04 &	 1.47 &	 FM &	 1.23 &	 FM &	 0.046 &	 Inv. \\
Cu$_2$ScFe & 	 -0.027 &	 0.172 &	 0.43 &	 1.49 &	 FM &	 0.57 &	 FM &	 0.048 &	 Reg. \\
Zn$_2$TiMn & 	 -0.131 &	 0.071 &	 0.26 &	 1.52 &	 FM &	 0.70 &	 FM &	 0.048 &	 Reg. \\
Rh$_2$MnFe & 	 -0.143 &	 0.000 &	 0.00 &	 1.33 &	 AFM &	 2.23 &	 FM &	 0.052 &	 Reg. \\
Fe$_2$HfPt & 	 -0.568 &	 0.010 &	 0.26 &	 1.48 &	 FM &	 1.26 &	 FM &	 0.057 &	 Inv. \\
Co$_2$VFe & 	 -0.109 &	 0.054 &	 0.78 &	 1.39 &	 FM &	 1.50 &	 FM &	 0.061 &	 Reg. \\
Co$_2$NbFe & 	 -0.054 &	 0.086 &	 0.79 &	 1.37 &	 FM &	 1.45 &	 FM &	 0.062 &	 Reg. \\
Fe$_2$VIr & 	 -0.209 &	 0.058 &	 0.00 &	 1.40 &	 AFM &	 1.01 &	 FM &	 0.070 &	 Reg. \\
Co$_2$TaFe & 	 -0.129 &	 0.086 &	 0.79 &	 1.36 &	 FM &	 1.40 &	 FM &	 0.073 &	 Reg. \\
Mn$_2$TiPt & 	 -0.387 &	 0.074 &	 0.15 &	 1.46 &	 AFM &	 0.42 &	 AFM &	 0.077 &	 Inv. \\
Mn$_2$CoNi & 	 -0.102 &	 0.000 &	 0.00 &	 1.40 &	 AFM &	 2.26 &	 FM &	 0.084 &	 Reg. \\
Mn$_2$ReIr & 	 -0.144 &	 0.000 &	 0.00 &	 1.37 &	 AFM &	 1.48 &	 FM &	 0.086 &	 Reg. \\
Cu$_2$TiMn & 	 -0.044 &	 0.112 &	 0.46 &	 1.55 &	 FM &	 0.98 &	 FM &	 0.093 &	 Reg. \\
Mn$_2$CoPt & 	 -0.256 &	 0.000 &	 0.00 &	 1.38 &	 AFM &	 2.26 &	 FM &	 0.095 &	 Reg. \\
Mn$_2$TaPt & 	 -0.279 &	 0.099 &	 0.00 &	 1.51 &	 AFM &	 1.23 &	 FM &	 0.108 &	 Inv. \\
Mn$_2$RhCo & 	 -0.158 &	 0.000 &	 0.00  &	 1.39 &	 AFM &	 2.03 &	 FM &	 0.110 &	 Reg. \\
Mn$_2$RhPt & 	 -0.383 &	 0.000 &	 0.00 &	 1.33 &	 AFM &	 2.26 &	 FM &	 0.116 &	 Reg. \\
Ni$_2$NbMn & 	 -0.158 &	 0.076 &	 0.58 &	 1.47 &	 FM &	 1.10 &	 FM &	 0.116 &	 Reg. \\
Ni$_2$TaMn & 	 -0.215 &	 0.087 &	 0.59 &	 1.46 &	 FM &	 1.06 &	 FM &	 0.119 &	 Reg. \\
Mn$_2$RhNi & 	 -0.217 &	 0.000 &	 0.00 &	 1.39 &	 AFM &	 2.24 &	 FM &	 0.127 &	 Reg. \\
Fe$_2$MoIr & 	 -0.058 &	 0.104 &	 0.88 &	 1.44 &	 FM &	 1.31 &	 FM &	 0.128 &	 Inv. \\
Fe$_2$CrIr & 	 -0.021 &	 0.066 &	 0.69 &	 1.43 &	 FM &	 1.65 &	 FM &	 0.136 &	 Inv. \\
Mn$_2$IrCo & 	 -0.206 &	 0.000 &	 0.00 &	 1.33 &	 AFM &	 2.00 &	 FM &	 0.142 &	 Reg. \\
Ni$_2$VMn & 	 -0.180 &	 0.035 &	 0.00 &	 1.46 &	 AFM &	 1.12 &	 FM &	 0.169 &	 Reg. \\ 
Mn$_2$FeIr & 	 -0.171 &	 0.000 &	 0.00 &	 1.41 &	 AFM &	 1.76 &	 FM &	 0.170 &	 Reg. \\
Fe$_2$ReIr & 	 -0.042 &	 0.025 &	 0.80 &	 1.43 &	 FM &	 1.44 &	 FM &	 0.201 &	 Inv. \\
Mn$_2$WIr & 	 -0.027 &	 0.126 &	 0.20 &	 1.48 &	 FM &	 1.20 &	 FM &	 0.205 &	 Inv. \\ 
 \bottomrule
	\end{tabular*}
	\label{Tab:TRAN}
\end{table*}

\subsection{Structural Transitions} 

Magneto-structural coupling is a key factor for achieving a large magnetization change and thus enhancing MCE and MSM applications, and enabling field control of the structural transition.  
The magneto-structural coupling can be quantified based on the concept of Curie temperature window (CTW), that has been previously applied to the experimental realization of a large MCE in the MM$^\prime$X (M,M$^\prime$ = metal, X = main group) family by disorder \cite{Wei2015b} and recently by us in the context of a computational workflow to search for novel MM$^\prime$X \cite{MMX2023}. 
In terms of CTW, the structural transitions should occur such that it involves two phases with different magnetization, {\it i.e.}, the structural phase transition temperature should lie in the temperature range spanned between the Curie temperatures of the martensite ($T^{mart}_C$) and austenite ($T^{aust}_C$) phases. 
Here the CTW concept has been extended in two ways. Firstly, the austenite may be the high magnetization phase during the transition ($T^{aust}_C>T^{mart}_C$), leading to an inverse MCE. Secondly, a magneto-structural transition can also occur between AFM and FM states, in which case the upper limit of the window is the T$_\textrm{C}$ of the FM phase. 

To implement the CTW concept for screening magnetocaloric materials, it is essential to estimate the martensite transition temperature ($T_M$), the magnetic ground states of the martensite and austenite and their respective magnetic ordering temperatures. The latter can be determined by using Monte Carlo simulations based on the exchange interactions calculated by DFT (more details in Methods). While the former requires an evaluation of the Gibbs free energy of the austenite and martensite phases, and determining the temperature where the austenite becomes lower in free energy. To account for the Gibbs free energies, we include the vibrational (F$^{\textrm{vib}}$), electronic (F$^{\textrm{el}}$) using the quasi-harmonic approximation (volume dependent) and magnetic F($^{\textrm{mag}}$) contributions , that can be expressed as 
\begin{equation}
 	F(T)=\mathop{min}_V\left\lbrace E_0(V)+F^{\textrm{vib}}(V,T)+F^{\textrm{el}}(V,T)\right\rbrace  + F^{\textrm{mag}}(T)
\end{equation}
  where $E_0$ are the T=0 energies from DFT \cite{Sozen2020}. Both F$^{vib}$(V,T) and F$^{el}$(V,T) contributions were calculated from DFT, based on phonons and the electronic density of states, as detailed in Methods. As for F$^{mag}$, we applied the modified Inden–Hillert–Jarl (IHJ) model \cite{Xiong2012}, using the T$_{\textrm{C/N}}$ from our CTW calculation and the DFT moments, to keep calculations tractable we exclude the volume dependency. 
 From the free energy, the T$_\textrm{M}$ can be deduced by finding at what temperature the austenite becomes the lowest energy structure.  
 
\begin{figure}
	\includegraphics[width=1.0\linewidth]{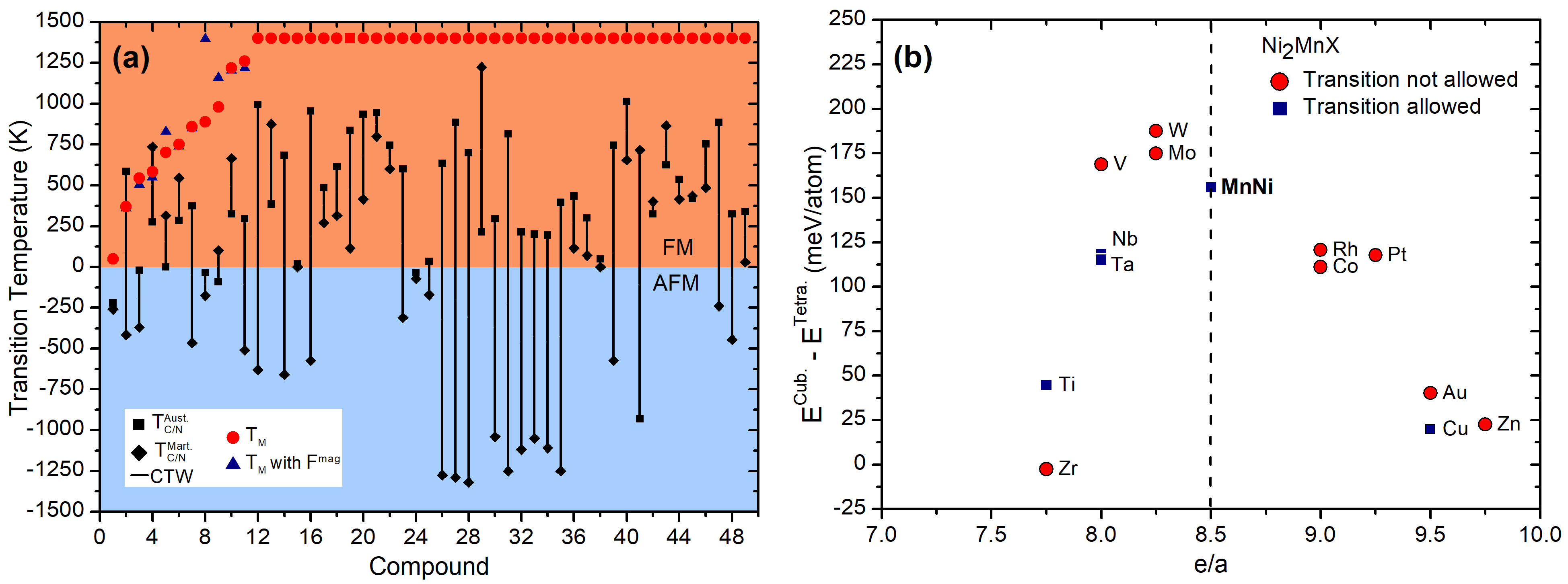}
	\caption{ (a) Plot of the CTW represented as error bars showing, with the structural transition from the free energy calculations represented by red square marks, with blue circles including the magnetic free energy. (b) The valence electron concentration ratio of all-d-metal Ni$_2$MnX compounds versus the energy difference in their hypothetical regular Heusler structure, this includes compounds that would otherwise be stable in the inverse Heusler or that have locally unstable austenite phases. The blue squares are compounds with allowed transition from the Bain path, while the centre line denotes the e/a value of MnNi.}
	\label{fig:transi}
\end{figure}

Figure \ref{fig:transi}(a) plots the T$_\textrm{M}$ (as red squares), where the compounds with 1400K denote that the martensite is always the lowest energy structure, hence there is no predicted transition. The calculated CTW is also represented as an error bar, in order to enable an easier comparison between values (c.f. Table S.7). We note that this represents the case in absence of substitutional disorder, which commonly use to tune both the magnetic and structural transition temperatures. On the whole we find 11 systems with a finite temperature structural transition, that we show in Table \ref{Tab:transition}.  Of all the free energy terms the F$^{vib}$ is the dominant term, with F$^{mag}$ usually having a small effect of a few dozen Kelvin on the structural transition temperature. The exception is Pd$_2$MnAg, where the martensite becomes more stable and the transition is no longer observed.
Noteworthy is the inclusion of Ni$_2$NbMn, since it shows a high Pugh ratio of 6.22. In this case, there is a wide CTW  due to  a high T$_\textrm{C}$ martensite and a lower T$_\textrm{C}$ austenite that is inline with a FM-PM transition, however the high T$_\textrm{M}$ precludes the possibility of magneto-structural transition without addition tuning, as we will discuss below.
In the case of Zn$_2$MnPd the transition lies squarely in the CTW, resulting in a predicted transition between an AFM martensite and FM austenite. Thus, the austenite is the high magnetization phase and the martensite the low magnetization state, similar to Ni-Mn-In compounds that exhibit an inverse MCE \cite{Gottschall2016}. In the case of Au$_2$MnCu the T$_\textrm{M}$ is too low and thermal arrest of the transition is expected.

\begin{table*}[]\centering
	\caption{Systems with a structural transition at finite temperatures, with the magnetic Néel/Curie temperatures (T$_{\textrm{C/N}}$) along with the T$_\textrm{M}$ with and without the F$^{\textrm{mag}}$ term. For mechanically stable phases the respective B/G ratios are also presented. }
		\vspace{0.1cm}
			\begin{tabular*}{0.9\linewidth}{@{\extracolsep{\fill}}ccccccc@{}}
				Compound   & Dist. convex hull & T$^{\textrm{mart}}_{\textrm{C/N}}$ & T$^{\textrm{aust}}_{\textrm{C/N}}$ & T$_\textrm{M}$ & Aust. B/G & Mart. B/G   \\
				\space     & [eV/atom]         & [K]                                & [K]                                & [K]            &           &             \\
				Au$_2$MnCu & 0.031             & -260                               & -220                               & 50/50          & 5.50      & 4.06        \\
				Zn$_2$MnPd & 0.078             & -415                               & 585                                & 360/370        & 3.67      & 2.50        \\
				Rh$_2$VMn  & 0.032             & -370                               & -20                                & 505/545        & ---       & 2.27        \\
				Cu$_2$ZrMn & 0.147             & 735                                & 275                                & 550/585        & 3.48      & 3.52        \\
				Zn$_2$HfMn & 0.121             & 315                                & 0                                  & 830/700        & 2.93      & 2.25        \\
				Cu$_2$ScFe & 0.172             & 545                                & 285                                & 740/750        & 8.13      & 3.29        \\
				Mn$_2$TiPt & 0.142             & -465                               & 375                                & 850/860        & 3.96      & 2.17        \\
				Pd$_2$MnAg & 0.034             & -175                               & -35                                & ---/890        & 4.19      & 3.22        \\
				Zn$_2$CrRh & 0.131             & 100                                & -90                                & 1160/980       & 2.50      & 2.44        \\
				Ni$_2$NbMn & 0.122             & 665                                & 325                                & 1205/1220      & 6.22      & 2.20        \\
				Fe$_2$VIr  & 0.127             & -510                               & 295                                & 1220/1260      & 3.82      & 1.70      
			\end{tabular*}
	\label{Tab:transition}
\end{table*}

\subsection{Substitutional tuning} 
Due to the stringent requirement of the T$_\textrm{M}$ falling inside the CTW, a large MCE is typically accomplished by lowering the T$_\textrm{M}$ or adjusting the CTW by substitutional disorder, as in the case of all-$d$-metal Ni-Co-Mn-Ti Heusler. Another strategy is isostructural substitution in the Mn(Ni,Co)(Ge,Si) MM'X alloy family, where one tunes towards a compound in the austenite phase (to control T$_\textrm{M}$) or towards a FM compound (to control the CTW) \cite{Wei2015b}. As previously mentioned, in Heusler alloys substitution is understood in terms of the number of valence electrons, where going way from the e/a of MnNi (8.5) lowers the transition temperature \cite{Wei2016}.
In the case of Ni$_2$Mn$_{1-x}$Ti$_x$, there is effectively an isostructural substitution of the NiMn alloy that has a high T$_\textrm{M}$ of 973 K \cite{Potapov} towards an austenite ground state, since the higher Ti content lowers T$_\textrm{M}$ until the transition vanishes. At the same time Co is added to manipulate the magnetic ground state and thus the CTW  \cite{Wei2015a,Taubel2020}.

 This e/a ratio dependence can also be seen in Ni$_2$MnX all-$d$-metal compounds, as observed in Figure \ref{fig:transi}(b) that shows e/a versus the energy difference of the austenite and martensite, under the assumption of a hypothetical regular Heusler structure. The compounds that are above or below the ratio of MnNi tend to have a lower energy difference between the phases (E$_0^{\textrm{Cub.}}$-E$_0^{\textrm{Tetra.}}$), which has been shown to relate well with the T$_\textrm{M}$ \cite{Barman2008}. However, the e/a ratio by itself does not account for the existence of a metastable austenite and stable martensite (compounds in blue squares). 
\par
\begin{figure}
	\includegraphics[width=\linewidth]{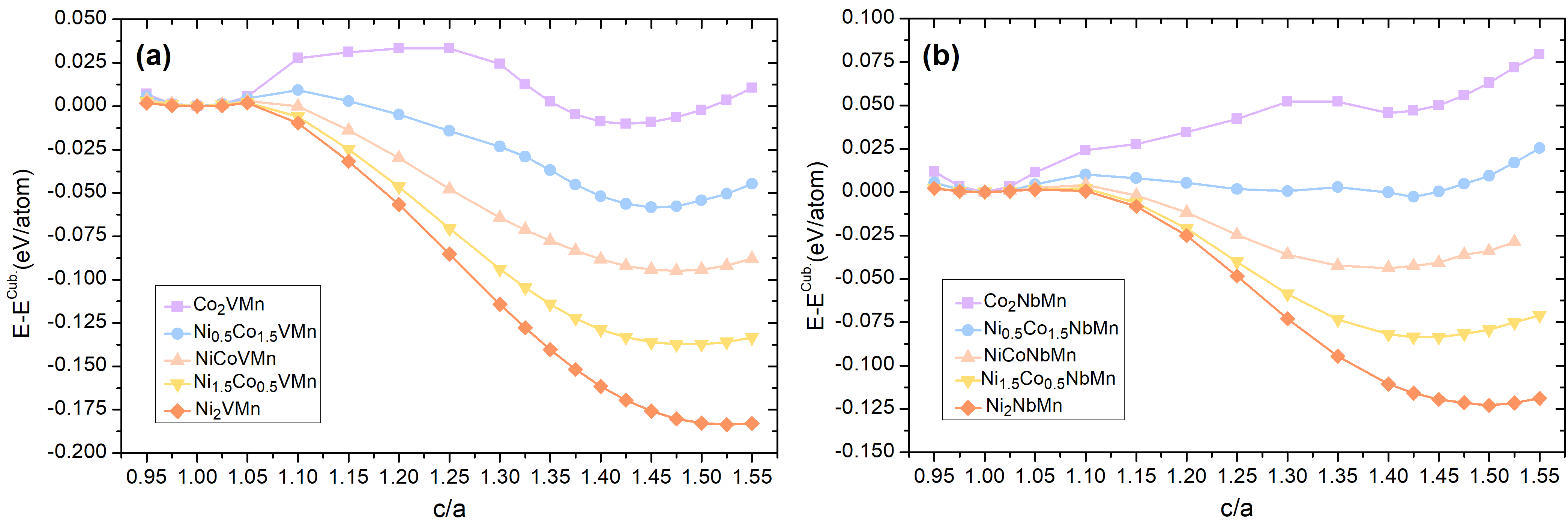}
	\caption{ (a) Bain Paths for Ni$_{2-x}$Co$_x$VMn based on EMTO-CPA results. The values of energy are given in respect to the energy of the cubic austenite at each composition to allow a straightforward comparison of the energy difference with tetragonal martensite. (b) The Bain path for Ni$_{2-x}$Co$_x$NbMn. }
	\label{fig:iso}
\end{figure}

As such, we propose to perform an isostructural substitution between a compound with an allowed transition and a compound stable in the cubic austenite (tetragonal martensite), to lower (increase) the E$_0^{\textrm{Cub.}}$-E$_0^{\textrm{Tetra.}}$, thus controlling T$_M$ as needed to establish magneto-structural coupling. This can be tested by calculating the Bain paths of compounds with substitutional disorder, to that end we adopted the EMTO code using the Coherent Potential Approximation (CPA) \cite{EMTO1,EMTO2,EMTO3,EMTO4,EMTO5,EMTO6,EMTO7}, that we have previously applied to explain the experimental finding of long-range ordering in Ni-Co-Mn-Ti systems \cite{Koch22}. For instance, Ni$_2$VMn has a large energy difference between both states (0.169 eV/atom), while Co$_2$VMn  is experimentally found in the austenite due to a small E$_0^{\textrm{Cub.}}$-E$_0^{\textrm{Tetra.}}$. Thus, the Ni$_{2-x}$Co$_x$VMn isostructural substitution allows for control of the Bain Path, as seen in the \ref{fig:iso}(a). At around Ni$_{0.5}$Co$_{1.5}$VMn (e/a=6.125) the energy difference fully stabilizes the martensite, while the austenite is meta-stable, with a further increase of Nickel content increasing the energy difference, and by expectation T$_\textrm{M}$ as desired. The same procedure can be used Ni$_2$NbMn by going towards the Co$_2$NbMn austenite phase, once again this allows control of the Bain path and the thus the transition, demonstrating isostructural substitution as method for tuning Heusler alloys and inducing magneto-structural coupling for MCE applications.

\section{Conclusions}\label{conclusions}
Based on HTP screening 	of all-$d$-metal Hesuler alloys we find 686 thermodynamically stable compounds, of which half is below 75 meV/atom above the convex Hull and the rest below 175 meV/atom. Based on our dataset we conclude that the inverse versus regular structural preference in all-$d$-metal Heusler alloys is driven by electronegativity difference between the A and B/C atoms, a low difference stabilizes the inverse structure,thus going against Burch's rule, which we tribute to metallic bonding. Likewise based on Hume-Rothery rules, we show that a large mismatch of radius of B/C elements compared the A element favours the formation a regular structure, where the A element sits in it's own sublattice. While Heusler alloys with main group elements prefer larger differences in electronegativity which favours an ionic bonding in the rock salt sublattice of the inverse Heusler. Following our HTP workflow for discovering materials with large MCE we establish possible structural transitions by calculating stable and metastable states in the form the FM and AFM Bain paths, we find 68 compounds with a double-well energy surface that allows or a structural transition, of these 49 are stable based on mechanical and dynamical stability criteria. Furthermore, by calculating finite-temperature magnetic behaviour and free energies of both martensite and austenite phases we narrow down to 11 systems with a possible finite-temperature structural transition. Based on the concept of isostructural substitution we provide a pathway to tune the transition in these novel Heusler systems.

\section*{Methods }
The HTP relaxation and total energy DFT calculations were performed using the Vienna Abinitio code (VASP) \cite{Kresse1996,Kresse1996a} within an in-house HTP code \cite{Singh2018}, the exchange correlation was treated using the PBE functional \cite{Perdew1996}. For the HTP screening the four atom primitive cell for the L2$_1$ and XA structures were used, along with the conventional unit cell of the I4/mmm and I$\overline{4}$m2 structures initialized at c/a=1.35 and then relaxed. The cut-off energy of the plane waves was set to 550 eV and for the k-mesh VASP’s automatic scheme was used, which generates a Gamma-centred Monkhorst-Pack grid with the k-mesh density  set to 35, i.e. the number of k-points in each direction is approximately 35 multiplied by the reciprocal lattice vector. The HTP AFM screening procedure was performed by considering different AFM configurations within a 8 atom cell for both cubic and tetragonal systems, that were re-relaxed in the AFM state. For systems with E$_{AFM}$$<$E$_{FM}$ we systematically perform the AFM Bain Path. The FM and AFM Bain paths were calculated using energy values from a Birch–Murnaghan fit of the energy-volume curve at a constant c/a ratio.
\par
The phonon stability and QHA vibrational free energies were obtained using the frozen phonon approach in phonopy \cite{Togo2015a} in 64 atom supercells, with the electronic contribution included using the fixed Density of States method, the volumes were calculated in 1.5$\%$ steps. Forces were calculated using VASP, with a k-mesh density of 50 and a energy cut-off of 540 eV. The single crystal elastic constants were calculated with a plane-wave cut-off of 700 eV, using VASP's built-in stress-strain method. Both calculations were carried out in the respective magnetic ground state of the structure, with a convergence criteria of 10$^{-7}$ eV. With the same settings the LOBSTER code \cite{Dronskowski1993,Maintz2016,Maintz2013} was used to calculate the COHP for selected compounds.  
\par
Magnetic exchange interactions \cite{Liech87} were calculated within a radius of 4 lattice parameters using the SPR-KKR code \cite{Ebert2011,Ebert} by applying the magnetic force theorem to the converged electronic structure of the magnetic ground state found earlier on. The angular moment cut-off of the spherical harmonics was set to $l$=3 and a k-mesh density was approximately 2000 points per irreducible Brillouin zone wedge. When electronic convergence was not achieved in SPRKKR, it was replaced by JuelichKKR code at comparable settings \cite{JuKKR}. 
The Currie temperature calculation by Monte Carlo Metropolis sampling, within a 10x10x10 supercell and the Curie temperature was determined by using the Binder cumulant in FM systems and heat capacity peak for AFM cases.
The Bain paths of selected disordered systems were calculated using the Coherent Potential Approximation as implemented in the EMTO-CPA code, which based on an all-electron Exact Muffin-Tin Orbitals formalism \cite{EMTO1,EMTO2,EMTO3,EMTO4,EMTO5,EMTO6,EMTO7}. The PBE functional was used along with soft-core approximation and 13x13x13 k-mesh. For the A-site ASA atomic sphere size was 1.08 that of the Wigner-Seitz spheres and the scale of potential spheres was chosen to be 0.95, while for other sites both values were set to 1.00.

\section*{Acknowledgements}\label{Acknowledgements}
We acknowledge the financial support from the European Research Council (ERC) under the European Union’s Horizon 2020 research and innovation programme (Grant No. 743116 - project” Cool Innov”) and the Deutsche Forschungsgemeinschaft (DFG, German Research Foundation) - Project-ID 405553726 - TRR 270. Lichtenberg high performance computer of the TU Darmstadt where calculations were performed for this project is gratefully acknowledged for the computational resources.

\bibliographystyle{unsrt}
\bibliography{library}

\newpage
\section*{ }\label{suplementary}


\section*{Supplementary material}

Table S.1 – The results of the HTP screening of all-d-metal Heuslers showing the formation energies, distance to the convex hull, lattice parameters and ground state and total magnetic moment. Note that the a lattice parameter is given in the 8-atom cell, and thus has a factor of $\sqrt{2}$ from the convention Heusler structure.

| Composition | Formation energy [ev/atom] | Dist. convex hull [ev/atom] | Lat. Param. a [Å] | Lat. Param. c [Å] | SPG | Mag. Ground State | Sum Mag. Mom. [muB/atom] |
|---|---|---|---|---|---|---|---|
| Fe2NiPt | -0.169 | 0.000 | 3.70 | 7.45 | 139 | FM | 1.65 |
| Fe2PtPd | -0.157 | 0.000 | 3.86 | 7.49 | 139 | FM | 1.66 |
| Ir2FeRe | -0.126 | 0.000 | 3.83 | 7.58 | 139 | FM | 0.52 |
| Mn2PdPt | -0.274 | 0.000 | 4.23 | 6.78 | 139 | FM | 2.13 |
| Mn2CrPt | -0.175 | 0.000 | 3.94 | 7.22 | 139 | FM | 1.01 |
| Pd2FeAu | -0.093 | 0.000 | 3.99 | 7.70 | 139 | FM | 0.94 |
| Pt2CoNi | -0.097 | 0.000 | 3.83 | 7.36 | 139 | FM | 0.84 |
| Pt2CrCu | -0.229 | 0.000 | 3.91 | 7.38 | 139 | FM | 0.62 |
| Pt2FeCo | -0.182 | 0.000 | 3.83 | 7.52 | 139 | FM | 1.40 |
| Pt2FeCr | -0.262 | 0.000 | 3.80 | 7.61 | 139 | FERI | 0.15 |
| Pt2FeCu | -0.230 | 0.000 | 3.93 | 7.23 | 139 | FM | 0.96 |
| Pt2FePd | -0.177 | 0.000 | 3.91 | 7.77 | 139 | FM | 0.98 |
| Pt2FeTa | -0.571 | 0.000 | 3.88 | 7.97 | 139 | FM | 0.59 |
| Pt2FeTi | -0.700 | 0.000 | 3.89 | 7.71 | 139 | FM | 0.63 |
| Pt2FeV | -0.455 | 0.000 | 3.83 | 7.68 | 139 | FM | 0.53 |
| Pt2MnNb | -0.537 | 0.000 | 3.92 | 7.99 | 139 | FM | 0.70 |
| Pt2MnTa | -0.585 | 0.000 | 3.92 | 7.97 | 139 | FM | 0.70 |
| Pt2MnTi | -0.714 | 0.000 | 3.97 | 7.62 | 139 | FM | 0.84 |
| Rh2CoRe | -0.023 | 0.000 | 3.78 | 7.53 | 139 | FM | 0.35 |
| Rh2FeRe | -0.065 | 0.000 | 3.80 | 7.58 | 139 | FM | 0.59 |
| Rh2MnRe | -0.094 | 0.000 | 3.82 | 7.57 | 139 | FM | 0.53 |
| Zn2MnRh | -0.249 | 0.000 | 3.86 | 7.55 | 139 | FM | 0.99 |
| Pt2MnCu | -0.309 | 0.000 | 3.96 | 7.31 | 139 | FM | 0.98 |
| Fe2ZrAu | -1.229 | 0.000 | 3.85 | 8.29 | 119 | FM | 1.03 |
| Fe2CrPt | -0.150 | 0.000 | 3.61 | 7.35 | 139 | FM | 0.63 |
| Mn2CoCr | -0.039 | 0.000 | 4.06 | 5.74 | 225 | FM | 1.22 |
| Nb2CoRe | -0.115 | 0.000 | 4.46 | 6.31 | 225 | FM | 0.42 |
| Rh2FeZn | -0.261 | 0.000 | 4.25 | 6.02 | 225 | FM | 1.07 |
| Rh2MnTi | -0.558 | 0.000 | 4.35 | 6.15 | 225 | FM | 1.17 |
| Rh2MnZn | -0.285 | 0.000 | 4.27 | 6.03 | 225 | FM | 0.81 |
| Ru2MnNb | -0.133 | 0.000 | 4.38 | 6.19 | 225 | FM | 0.99 |
| Ru2MnTa | -0.241 | 0.000 | 4.38 | 6.20 | 225 | FM | 0.99 |
| Ta2CoRe | -0.248 | 0.000 | 4.46 | 6.30 | 225 | FM | 0.38 |
| Ta2FeOs | -0.282 | 0.000 | 4.45 | 6.29 | 225 | FM | 0.43 |
| Ti2CoIr | -0.648 | 0.000 | 4.33 | 6.13 | 225 | FM | 0.40 |
| Zr2FePd | -0.520 | 0.000 | 4.63 | 6.55 | 225 | FM | 0.16 |
| Co2FeMn | -0.039 | 0.000 | 4.04 | 5.72 | 225 | FM | 2.26 |
| Co2TiMn | -0.275 | 0.000 | 4.13 | 5.84 | 225 | FM | 1.26 |
| Fe2IrPt | -0.156 | 0.000 | 3.89 | 7.19 | 139 | AFM | 0.00 |
| Fe2RhIr | -0.088 | 0.000 | 3.88 | 7.02 | 139 | AFM | 0.00 |
| Fe2RhPt | -0.164 | 0.000 | 3.90 | 7.16 | 139 | AFM | 0.00 |
| Ir2MnFe | -0.186 | 0.000 | 3.77 | 7.10 | 139 | AFM | 0.00 |
| Ir2MnRu | -0.137 | 0.000 | 3.80 | 7.64 | 139 | AFM | 0.00 |
| Mn2AuPt | -0.255 | 0.000 | 4.71 | 5.62 | 139 | AFM | 0.00 |
| Mn2IrOs | -0.176 | 0.000 | 3.81 | 7.32 | 139 | AFM | 0.00 |
| Mn2IrPt | -0.937 | 0.000 | 3.90 | 7.35 | 139 | AFM | 0.00 |
| Mn2IrRu | -0.193 | 0.000 | 3.81 | 7.31 | 139 | AFM | 0.00 |

| | | | | | | | |
|---|---|---|---|---|---|---|---|
| Mn2NiPd | -0.203 | 0.000 | 3.79 | 7.38 | 139 | AFM | 0.00 |
| Mn2PtOs | -0.191 | 0.000 | 3.87 | 7.27 | 139 | AFM | 0.00 |
| Mn2PtRu | -0.243 | 0.000 | 3.87 | 7.28 | 139 | AFM | 0.00 |
| Mn2RhIr | -0.303 | 0.000 | 3.84 | 7.26 | 139 | AFM | 0.00 |
| Mn2RuRh | -0.130 | 0.000 | 3.81 | 7.26 | 139 | AFM | 0.00 |
| Ni2MnZn | -0.186 | 0.000 | 3.77 | 6.79 | 139 | AFM | 0.00 |
| Pd2MnCd | -0.302 | 0.000 | 4.20 | 7.37 | 139 | AFM | 0.00 |
| Pd2MnHg | -0.201 | 0.000 | 4.17 | 7.58 | 139 | AFM | 0.00 |
| Pd2MnZn | -0.424 | 0.000 | 4.07 | 7.07 | 139 | AFM | 0.00 |
| Pt2CrMn | -0.567 | 0.000 | 3.98 | 7.29 | 139 | AFM | 0.00 |
| Pt2CrPd | -0.216 | 0.000 | 3.90 | 7.84 | 139 | AFM | 0.00 |
| Pt2CrZn | -0.392 | 0.000 | 3.96 | 7.44 | 139 | AFM | 0.00 |
| Pt2NiCr | -0.240 | 0.000 | 3.83 | 7.48 | 139 | AFM | 0.00 |
| Pt2VCr | -0.422 | 0.000 | 3.84 | 7.74 | 139 | AFM | 0.00 |
| Pt2MnCd | -0.337 | 0.000 | 4.14 | 7.57 | 139 | AFM | 0.00 |
| Pt2MnNi | -0.299 | 0.000 | 3.87 | 7.43 | 139 | AFM | 0.00 |
| Pt2MnV | -0.527 | 0.000 | 3.88 | 7.65 | 139 | AFM | 0.00 |
| Pt2MnZn | -0.519 | 0.000 | 4.04 | 7.19 | 139 | AFM | 0.00 |
| Rh2FeCr | -0.116 | 0.000 | 3.81 | 7.26 | 139 | AFM | 0.00 |
| Rh2NiMn | -0.107 | 0.000 | 3.79 | 7.25 | 139 | AFM | 0.00 |
| Rh2MnIr | -0.150 | 0.000 | 3.81 | 7.67 | 139 | AFM | 0.00 |
| Ir2MnZn | -0.195 | 0.000 | 3.93 | 7.13 | 139 | AFM | 0.00 |
| Mn2CoIr | -0.206 | 0.000 | 3.74 | 7.05 | 139 | AFM | 0.00 |
| Mn2CoNi | -0.102 | 0.000 | 3.58 | 7.09 | 139 | AFM | 0.00 |
| Mn2CoRh | -0.158 | 0.000 | 3.69 | 7.23 | 139 | AFM | 0.00 |
| Mn2FeIr | -0.171 | 0.000 | 3.66 | 7.30 | 139 | AFM | 0.00 |
| Mn2FePt | -0.207 | 0.000 | 3.73 | 7.29 | 139 | AFM | 0.00 |
| Mn2MoPt | -0.625 | 0.000 | 3.89 | 7.38 | 139 | AFM | 0.00 |
| Mn2PtCo | -0.256 | 0.000 | 3.75 | 7.33 | 139 | AFM | 0.00 |
| Mn2PtRh | -0.383 | 0.000 | 3.90 | 7.34 | 139 | AFM | 0.00 |
| Mn2ReIr | -0.144 | 0.000 | 3.80 | 7.38 | 139 | AFM | 0.00 |
| Mn2NiRh | -0.217 | 0.000 | 3.72 | 7.31 | 139 | AFM | 0.00 |
| Rh2MnFe | -0.143 | 0.000 | 3.84 | 7.22 | 139 | AFM | 0.00 |
| Hg2MnY | -0.143 | 0.000 | 5.01 | 7.08 | 225 | AFM | 0.00 |
| Ir2CrZn | -0.146 | 0.000 | 4.28 | 6.05 | 225 | AFM | 0.00 |
| Rh2CrZn | -0.209 | 0.000 | 4.26 | 6.03 | 225 | AFM | 0.00 |
| Ru2CrSc | -0.133 | 0.000 | 4.37 | 6.18 | 225 | AFM | 0.00 |
| Ru2CrTi | -0.235 | 0.000 | 4.30 | 6.08 | 225 | AFM | 0.00 |
| Ru2MnSc | -0.158 | 0.000 | 4.38 | 6.20 | 225 | AFM | 0.00 |
| Sc2MnOs | -0.199 | 0.000 | 4.62 | 6.53 | 225 | AFM | 0.00 |
| Sc2MnRu | -0.224 | 0.000 | 4.61 | 6.53 | 225 | AFM | 0.00 |
| Ti2MnMo | -0.091 | 0.000 | 4.38 | 6.19 | 225 | AFM | 0.00 |
| Y2MnPd | -0.269 | 0.000 | 5.00 | 7.07 | 225 | AFM | 0.00 |
| Co2NbHf | -0.197 | 0.000 | 4.40 | 6.23 | 225 | AFM | 0.00 |
| Co2NbTi | -0.192 | 0.000 | 4.28 | 6.06 | 225 | AFM | 0.00 |
| Co2TaHf | -0.235 | 0.000 | 4.40 | 6.22 | 225 | AFM | 0.00 |
| Co2TaTi | -0.246 | 0.000 | 4.28 | 6.06 | 225 | AFM | 0.00 |
| Co2ZrNb | -0.141 | 0.000 | 4.43 | 6.26 | 225 | AFM | 0.00 |
| Co2ZrTa | -0.175 | 0.000 | 4.42 | 6.25 | 225 | AFM | 0.00 |
| Au2MnPd | -0.196 | 0.000 | 4.02 | 8.12 | 119 | FM | 1.05 |
| Ni2FePt | -0.160 | 0.000 | 3.69 | 7.35 | 119 | FM | 1.13 |
| Pd2FePt | -0.143 | 0.000 | 3.91 | 7.73 | 119 | FM | 0.99 |
| Co2ZnPd | -1.412 | 0.000 | 3.78 | 7.16 | 119 | FM | 0.87 |

| | | | | | | | |
|---|---|---|---|---|---|---|---|
| Mn2NiFe | -0.060 | 0.000 | 3.71 | 6.68 | 119 | FM | 0.59 |
| Fe2MnIr | -9.442 | 0.000 | 3.68 | 7.15 | 119 | AFM | 0.00 |
| Au2CrPd | -0.074 | 0.000 | 4.00 | 8.20 | 119 | AFM | 0.00 |
| Cu2MnPd | -0.108 | 0.000 | 3.79 | 7.47 | 119 | AFM | 0.00 |
| Cu2MnPt | -0.200 | 0.000 | 3.81 | 7.40 | 119 | AFM | 0.00 |
| Ir2FeRh | -0.066 | 0.000 | 3.81 | 7.59 | 119 | AFM | 0.00 |
| Ir2MnRh | -0.163 | 0.000 | 3.81 | 7.66 | 119 | AFM | 0.00 |
| Mn2CuPt | -0.250 | 0.000 | 3.76 | 7.72 | 119 | AFM | 0.03 |
| Mn2RhZn | -0.209 | 0.000 | 3.78 | 7.60 | 119 | AFM | 0.04 |
| Mn2ZnPt | -0.659 | 0.000 | 3.83 | 7.64 | 119 | AFM | 0.04 |
| Ni2CrPt | -0.150 | 0.000 | 3.66 | 7.45 | 119 | AFM | 0.00 |
| Ni2MnPt | -0.210 | 0.000 | 3.68 | 7.44 | 119 | AFM | 0.00 |
| Pd2CrAu | -0.087 | 0.000 | 3.95 | 8.05 | 119 | AFM | 0.00 |
| Pd2CrPt | -0.147 | 0.000 | 3.90 | 7.82 | 119 | AFM | 0.00 |
| Pd2MnAu | -0.237 | 0.000 | 3.96 | 8.02 | 119 | AFM | 0.00 |
| Re2MnPt | -1.536 | 0.000 | 3.89 | 7.55 | 119 | AFM | 0.00 |
| Rh2MnHf | -0.654 | 0.000 | 4.47 | 6.32 | 225 | FM | 1.15 |
| Pd2MnCu | -0.215 | 0.000 | 4.33 | 6.12 | 225 | FM | 1.09 |
| Rh2MnSc | -0.594 | 0.000 | 4.44 | 6.27 | 225 | FM | 1.06 |
| Ru2FeTa | -0.180 | 0.000 | 4.37 | 6.18 | 225 | FM | 1.14 |
| Pt2NiFe | -0.210 | 0.002 | 3.86 | 7.36 | 139 | FM | 1.08 |
| Zn2MnIr | -0.190 | 0.002 | 3.86 | 7.56 | 139 | FM | 0.97 |
| Ni2MnRh | -0.096 | 0.003 | 3.65 | 7.31 | 119 | AFM | 0.00 |
| Os2MnTa | -0.151 | 0.003 | 4.40 | 6.23 | 225 | FM | 0.95 |
| Au2MnZn | -0.146 | 0.003 | 4.52 | 6.39 | 225 | FM | 0.97 |
| Co2HfZr | -0.344 | 0.003 | 4.48 | 6.34 | 225 | FM | 0.32 |
| Ru2VMn | -0.141 | 0.004 | 4.25 | 6.01 | 225 | FM | 0.97 |
| Pt2MnRh | -0.261 | 0.005 | 3.88 | 7.82 | 139 | AFM | 0.00 |
| Ir2MnCo | -0.106 | 0.005 | 3.70 | 7.57 | 119 | FM | 0.62 |
| Rh2MnZr | -0.572 | 0.005 | 4.49 | 6.36 | 225 | FM | 1.15 |
| Rh2MnMo | -0.170 | 0.006 | 3.86 | 7.57 | 139 | AFM | 0.00 |
| Rh2MnV | -0.307 | 0.006 | 3.87 | 7.24 | 139 | AFM | 0.00 |
| Ir2MnTi | -0.570 | 0.006 | 4.36 | 6.17 | 225 | FM | 1.17 |
| Pt2CoTi | -0.680 | 0.007 | 3.86 | 7.69 | 139 | FM | 0.40 |
| Ni2CoPt | -0.073 | 0.008 | 3.65 | 7.33 | 119 | FM | 0.89 |
| Pt2FeNb | -0.512 | 0.008 | 3.89 | 7.98 | 139 | FM | 0.60 |
| Ir2FeCo | -0.041 | 0.008 | 3.70 | 7.53 | 119 | FM | 0.89 |
| Co2ReIr | -0.042 | 0.009 | 3.69 | 7.45 | 119 | FM | 0.43 |
| Pt2NiPd | -0.040 | 0.009 | 3.87 | 7.73 | 139 | FM | 0.39 |
| Fe2HfPt | -0.568 | 0.010 | 3.80 | 7.95 | 119 | FM | 1.11 |
| Rh2FeNi | -0.030 | 0.010 | 3.80 | 7.11 | 139 | AFM | 0.00 |
| Ir2FeZn | -0.104 | 0.010 | 4.27 | 6.04 | 225 | FM | 1.03 |
| Ta2FeRu | -0.238 | 0.011 | 4.43 | 6.26 | 225 | FM | 0.41 |
| Rh2MnRu | -0.071 | 0.011 | 3.79 | 7.63 | 139 | AFM | 0.00 |
| Ni2CoFe | -0.057 | 0.011 | 3.51 | 7.22 | 139 | FM | 1.45 |
| Mn2CrRh | -0.071 | 0.011 | 4.24 | 5.99 | 225 | FM | 1.20 |
| Co2VPt | -0.285 | 0.011 | 3.67 | 7.42 | 119 | FM | 0.46 |
| Mn2RhPd | -0.166 | 0.012 | 3.91 | 7.35 | 139 | AFM | 0.00 |
| Mn2CoSc | -0.169 | 0.012 | 4.26 | 6.02 | 119 | FERI | 0.50 |
| Fe2CoTi | -0.229 | 0.013 | 4.10 | 5.80 | 216 | FM | 1.27 |
| V2FeOs | -0.232 | 0.013 | 4.20 | 5.94 | 225 | FM | 0.43 |
| Hf2MnRe | -0.240 | 0.013 | 4.59 | 6.49 | 225 | FM | 0.40 |
| Pt2FeRh | -0.144 | 0.013 | 3.86 | 7.80 | 139 | FM | 1.11 |

| | | | | | | |
|---|---|---|---|---|---|---|
| Pt2MnSc | -0.809 | 0.014 | 4.18 | 7.46 | 139 | FM | 1.02 |
| Ir2FeW | -0.238 | 0.014 | 3.88 | 7.56 | 139 | FM | 0.58 |
| Ni2FeRh | -0.050 | 0.014 | 3.62 | 7.40 | 139 | FM | 1.29 |
| Au2MnPt | -0.169 | 0.015 | 4.06 | 7.99 | 119 | AFM | 0.00 |
| Co2TaPt | -0.386 | 0.017 | 3.76 | 7.71 | 119 | FM | 0.47 |
| Ir2MnW | -0.251 | 0.018 | 3.88 | 7.59 | 139 | FM | 0.42 |
| Ir2MnOs | -0.093 | 0.018 | 3.81 | 7.66 | 139 | FM | 0.48 |
| Pt2CrRh | -0.192 | 0.018 | 3.86 | 7.79 | 119 | AFM | 0.00 |
| Fe2NiRh | -0.031 | 0.019 | 3.70 | 7.22 | 139 | FM | 1.70 |
| Ni2PtPd | -0.042 | 0.019 | 3.84 | 7.24 | 139 | FM | 0.54 |
| Pt2MnPd | -0.287 | 0.021 | 3.92 | 7.87 | 139 | AFM | 0.00 |
| Rh2MnCu | -0.061 | 0.021 | 3.85 | 7.19 | 139 | AFM | 0.00 |
| Co2HfTi | -0.371 | 0.022 | 4.34 | 6.14 | 225 | FM | 0.39 |
| Pt2CoRh | -0.033 | 0.022 | 3.84 | 7.73 | 139 | FM | 0.75 |
| Fe2CrNi | -0.011 | 0.023 | 3.57 | 7.15 | 139 | FM | 0.64 |
| Co2HfMn | -0.259 | 0.023 | 4.27 | 6.04 | 225 | FM | 1.27 |
| Fe2VPt | -0.271 | 0.024 | 3.70 | 7.62 | 119 | FM | 0.89 |
| Co2FeNi | -0.023 | 0.024 | 3.55 | 7.05 | 119 | FM | 1.63 |
| Pd2CrCu | -0.061 | 0.024 | 3.90 | 7.43 | 139 | AFM | 0.00 |
| Mn2PtNi | -0.181 | 0.024 | 3.79 | 7.40 | 139 | AFM | 0.00 |
| Ir2VMn | -0.374 | 0.025 | 3.84 | 7.34 | 139 | FM | 0.39 |
| Co2NiPt | -0.062 | 0.025 | 3.63 | 7.43 | 139 | FM | 1.18 |
| Fe2ReIr | -0.042 | 0.025 | 3.71 | 7.52 | 119 | FM | 0.82 |
| Ni2VFe | -0.185 | 0.025 | 3.52 | 7.39 | 139 | FM | 0.48 |
| Co2PtTi | -0.455 | 0.026 | 3.74 | 7.52 | 119 | FM | 0.57 |
| Cu2MnAu | -0.020 | 0.026 | 3.85 | 7.61 | 139 | FM | 0.98 |
| Fe2NiPd | -0.042 | 0.026 | 3.68 | 7.46 | 139 | FM | 1.65 |
| Pd2MnV | -0.242 | 0.026 | 3.88 | 7.66 | 139 | AFM | 0.00 |
| Co2FePt | -0.093 | 0.026 | 3.68 | 7.38 | 119 | FM | 1.62 |
| Fe2VRu | -0.099 | 0.027 | 4.15 | 5.87 | 216 | FM | 1.24 |
| Ir2MnHf | -0.624 | 0.027 | 4.48 | 6.34 | 225 | FM | 1.18 |
| Fe2CoRh | -0.017 | 0.027 | 4.15 | 5.87 | 225 | FM | 2.18 |
| Pt2FeZn | -0.368 | 0.028 | 3.98 | 7.28 | 139 | FM | 0.94 |
| Mn2VPt | -0.292 | 0.028 | 4.29 | 6.07 | 225 | FM | 1.24 |
| Fe2PdRh | -0.025 | 0.028 | 3.88 | 7.25 | 139 | FM | 1.73 |
| Co2CrIr | -0.058 | 0.029 | 3.65 | 7.23 | 119 | FERI | 0.28 |
| Au2MnCu | -0.059 | 0.029 | 4.18 | 7.21 | 139 | FM | 0.99 |
| Fe2CrRh | -0.014 | 0.029 | 3.69 | 7.42 | 139 | FM | 0.72 |
| Fe2TaRu | -0.140 | 0.029 | 4.28 | 6.06 | 216 | FM | 1.25 |
| Ni2MnCu | -0.041 | 0.030 | 3.67 | 6.93 | 139 | FM | 1.02 |
| Ni2PtRh | -0.031 | 0.031 | 3.71 | 7.60 | 119 | FM | 0.60 |
| Rh2FeW | -0.158 | 0.031 | 3.85 | 7.53 | 139 | FM | 0.61 |
| Co2PtNb | -0.307 | 0.031 | 3.75 | 7.72 | 119 | FM | 0.46 |
| Rh2MnW | -0.152 | 0.031 | 3.88 | 7.53 | 139 | FM | 0.69 |
| Cu2MnNi | -0.004 | 0.031 | 3.68 | 7.16 | 119 | AFM | 0.00 |
| Ni2FeCu | -0.029 | 0.032 | 3.63 | 6.91 | 139 | FM | 1.00 |
| V2FeRu | -0.177 | 0.032 | 4.18 | 5.91 | 225 | FM | 0.41 |
| Rh2CoPt | -0.012 | 0.032 | 3.81 | 7.59 | 119 | FM | 0.62 |
| Co2MnIr | -0.042 | 0.032 | 3.61 | 7.27 | 119 | AFM | 0.00 |
| Ru2FeV | -0.085 | 0.032 | 4.23 | 5.99 | 225 | FM | 1.10 |
| Pd2NiFe | -0.066 | 0.033 | 3.85 | 7.34 | 139 | FM | 1.07 |
| Pd2MnTa | -0.066 | 0.033 | 3.90 | 8.03 | 139 | FM | 0.68 |
| Pt2MnW | -0.263 | 0.033 | 3.86 | 7.97 | 139 | FM | 0.60 |

| | | | | | | |
|---|---|---|---|---|---|---|
| Ir2FeOs | -0.012 | 0.033 | 3.80 | 7.65 | 139 | FM | 0.59 |
| Co2TiFe | -0.217 | 0.033 | 4.12 | 5.82 | 225 | FM | 1.36 |
| Ru2FeNb | -0.069 | 0.034 | 4.36 | 6.17 | 225 | FM | 1.14 |
| Mn2IrCr | -0.098 | 0.034 | 4.24 | 5.99 | 225 | FM | 1.20 |
| Pd2MnAg | -0.146 | 0.034 | 4.14 | 7.38 | 139 | FM | 1.07 |
| Nb2FeOs | -0.141 | 0.034 | 4.45 | 6.29 | 225 | FM | 0.46 |
| Pd2NiPt | -0.014 | 0.035 | 3.86 | 7.72 | 119 | FM | 0.44 |
| Ni2MnW | -0.037 | 0.035 | 3.60 | 7.56 | 139 | FM | 0.50 |
| Pt2CoCr | -0.138 | 0.035 | 3.79 | 7.51 | 139 | AFM | 0.00 |
| Ru2MnIr | -0.050 | 0.035 | 3.78 | 7.63 | 119 | AFM | 0.00 |
| Ni2MnV | -0.180 | 0.035 | 3.54 | 7.42 | 139 | FM | 0.55 |
| Rh2FeMo | -0.132 | 0.036 | 3.85 | 7.52 | 139 | FM | 0.62 |
| Co2ZrTi | -0.306 | 0.036 | 4.37 | 6.18 | 225 | FM | 0.42 |
| Ir2MnMo | -0.233 | 0.036 | 3.87 | 7.58 | 139 | FM | 0.43 |
| Pt2CrCd | -0.201 | 0.036 | 4.07 | 7.79 | 139 | AFM | 0.00 |
| Ni2FeZn | -0.125 | 0.036 | 3.71 | 6.87 | 139 | FM | 0.98 |
| Os2VMn | -0.108 | 0.037 | 4.26 | 6.03 | 225 | FM | 0.89 |
| Pt2MnAu | -0.232 | 0.037 | 3.99 | 7.95 | 139 | AFM | 0.00 |
| Pt2MnHf | -0.804 | 0.037 | 3.93 | 8.39 | 119 | FM | 0.90 |
| Au2MnRh | -0.058 | 0.037 | 4.00 | 8.00 | 119 | AFM | 0.00 |
| Au2CrPt | -0.048 | 0.037 | 4.01 | 8.16 | 119 | AFM | 0.00 |
| Pd2FeV | -0.174 | 0.038 | 3.80 | 7.75 | 139 | FM | 0.45 |
| Ir2FeMo | -0.209 | 0.038 | 3.86 | 7.57 | 139 | FM | 0.56 |
| Ni2MnMo | -0.033 | 0.038 | 3.59 | 7.57 | 139 | FM | 0.50 |
| Fe2CoIr | -0.019 | 0.039 | 3.70 | 7.05 | 139 | AFM | 0.00 |
| Rh2CrIr | -0.117 | 0.039 | 3.79 | 7.67 | 139 | AFM | 0.00 |
| Au2NiMn | -0.061 | 0.040 | 3.89 | 8.02 | 119 | FM | 1.06 |
| Pd2NiMn | -0.141 | 0.040 | 3.88 | 7.38 | 139 | AFM | 0.00 |
| Ir2VFe | -0.350 | 0.040 | 3.84 | 7.31 | 139 | FM | 0.53 |
| Fe2PtW | -0.095 | 0.041 | 3.75 | 7.66 | 119 | FM | 0.81 |
| Ir2VCr | -0.349 | 0.041 | 3.83 | 7.41 | 139 | AFM | 0.00 |
| Pt2CoCu | -0.085 | 0.042 | 3.87 | 7.34 | 139 | FM | 0.65 |
| Mn2WIr | -0.111 | 0.042 | 3.73 | 7.82 | 119 | FERI | 0.13 |
| Fe2PtTa | -0.322 | 0.043 | 3.77 | 7.91 | 119 | FM | 0.96 |
| Co2VNi | -0.158 | 0.043 | 3.53 | 7.17 | 119 | FM | 0.44 |
| Fe2OsV | -0.086 | 0.043 | 4.17 | 5.89 | 216 | FM | 1.24 |
| Mn2MoRh | -0.050 | 0.043 | 4.27 | 6.04 | 225 | FM | 1.28 |
| Co2VIr | -0.245 | 0.043 | 3.66 | 7.31 | 119 | FM | 0.47 |
| Pt2FeMn | -0.243 | 0.044 | 3.96 | 7.38 | 139 | FM | 1.82 |
| Ti2CrOs | -0.358 | 0.044 | 4.36 | 6.16 | 225 | FM | 0.39 |
| Fe2IrZn | -0.050 | 0.044 | 4.20 | 5.94 | 216 | FM | 1.44 |
| Co2WIr | -0.128 | 0.045 | 3.72 | 7.49 | 119 | FM | 0.45 |
| Ni2FePd | -0.049 | 0.045 | 3.67 | 7.36 | 119 | FM | 1.13 |
| Pt2MnZr | -0.757 | 0.045 | 3.95 | 8.41 | 119 | FM | 0.90 |
| Au2MnCd | -0.087 | 0.045 | 4.67 | 6.61 | 225 | FM | 1.00 |
| Mn2RhRe | -0.037 | 0.045 | 3.73 | 7.53 | 119 | FERI | 0.01 |
| Rh2FeV | -0.264 | 0.045 | 3.85 | 7.22 | 139 | FM | 0.64 |
| Fe2TiIr | -0.382 | 0.045 | 4.24 | 5.99 | 216 | FM | 1.25 |
| Ir2CoW | -0.217 | 0.045 | 3.86 | 7.54 | 139 | FM | 0.36 |
| Fe2CoPt | -0.099 | 0.046 | 3.73 | 7.33 | 139 | FM | 1.86 |
| Rh2MnCd | -0.035 | 0.047 | 4.49 | 6.14 | 139 | AFM | 0.00 |
| Zn2CoNi | -0.107 | 0.047 | 3.85 | 7.02 | 139 | FM | 0.88 |
| V2CoNi | -0.165 | 0.048 | 3.65 | 7.31 | 139 | NM | 0.00 |

| | | | | | | |
|---|---|---|---|---|---|---|
| Zn2MnRu | -0.032 | 0.048 | 3.81 | 7.66 | 139 | FM | 1.02 |
| Mn2HfTi | -0.175 | 0.049 | 4.36 | 6.17 | 225 | FM | 0.40 |
| Co2PtRh | -0.020 | 0.049 | 3.80 | 7.31 | 139 | FM | 1.23 |
| Pt2CrW | -0.204 | 0.049 | 3.85 | 8.01 | 139 | AFM | 0.00 |
| Au2MnAg | -0.038 | 0.049 | 4.06 | 8.33 | 119 | FM | 1.02 |
| Pt2FeW | -0.203 | 0.050 | 3.82 | 7.97 | 139 | FM | 0.52 |
| Nb2FeRu | -0.105 | 0.051 | 4.43 | 6.27 | 225 | FM | 0.45 |
| Zn2MnAu | -0.084 | 0.051 | 4.10 | 7.29 | 119 | FM | 0.89 |
| Mn2PdAu | -0.095 | 0.051 | 4.48 | 6.34 | 225 | FM | 2.05 |
| Fe2TaOs | -0.106 | 0.051 | 4.31 | 6.09 | 216 | FM | 1.26 |
| Pt2FeMo | -0.226 | 0.051 | 3.83 | 7.96 | 139 | FM | 0.53 |
| Fe2NbTa | -0.105 | 0.052 | 4.34 | 6.13 | 225 | FM | 0.53 |
| Rh2MnOs | -0.029 | 0.053 | 3.79 | 7.65 | 139 | FM | 0.58 |
| Fe2NbPt | -0.259 | 0.053 | 3.77 | 7.93 | 119 | FM | 0.97 |
| Fe2NbRu | -0.050 | 0.053 | 4.28 | 6.06 | 216 | FM | 1.26 |
| Ni2FeW | -0.022 | 0.054 | 3.56 | 7.57 | 139 | FM | 0.36 |
| Cu2CrPt | -0.067 | 0.054 | 3.76 | 7.59 | 119 | AFM | 0.00 |
| Co2FeV | -0.109 | 0.054 | 3.58 | 7.06 | 139 | FM | 0.81 |
| Co2MnZr | -0.190 | 0.054 | 4.30 | 6.08 | 225 | FM | 1.29 |
| Nb2CoPt | -0.337 | 0.054 | 3.93 | 7.91 | 139 | NM | 0.00 |
| Pt2CoHf | -0.786 | 0.055 | 3.94 | 7.99 | 139 | FM | 0.38 |
| Ta2FePt | -0.313 | 0.055 | 4.01 | 7.66 | 139 | NM | 0.00 |
| Mn2PtCd | -0.114 | 0.055 | 3.96 | 8.06 | 119 | AFM | 0.03 |
| Mn2VRh | -0.173 | 0.055 | 4.23 | 5.99 | 225 | FM | 1.09 |
| Au2FePd | -0.011 | 0.056 | 3.96 | 8.13 | 119 | FM | 0.89 |
| Fe2VNi | -0.116 | 0.056 | 3.55 | 7.39 | 119 | FM | 0.88 |
| Fe2CoNi | -0.010 | 0.056 | 3.63 | 6.90 | 139 | FM | 1.85 |
| Cu2MnRh | -0.026 | 0.056 | 3.79 | 7.28 | 119 | FM | 0.95 |
| Pt2CoMn | -0.206 | 0.056 | 3.87 | 7.53 | 139 | FM | 1.46 |
| Mn2VIr | -0.224 | 0.057 | 4.23 | 5.99 | 225 | FM | 1.05 |
| Cu2FePt | -0.075 | 0.057 | 3.76 | 7.41 | 119 | FM | 0.85 |
| Ni2FeTa | -0.241 | 0.057 | 3.64 | 7.58 | 139 | FM | 0.48 |
| Co2WPt | -0.112 | 0.057 | 3.70 | 7.64 | 119 | FM | 0.38 |
| Ir2MnNb | -0.374 | 0.057 | 4.00 | 7.48 | 139 | FM | 0.77 |
| Rh2FeSc | -0.487 | 0.057 | 4.41 | 6.23 | 225 | FM | 1.09 |
| Fe2PtCu | -0.067 | 0.058 | 3.69 | 7.68 | 139 | FM | 1.52 |
| Ni2FeMo | -0.006 | 0.058 | 3.56 | 7.58 | 139 | FM | 0.40 |
| Pt2FeHf | -0.783 | 0.058 | 3.97 | 7.99 | 139 | FM | 0.63 |
| Pd2FeRh | -0.017 | 0.058 | 3.84 | 7.86 | 139 | FM | 1.22 |
| Fe2VIr | -0.209 | 0.058 | 3.70 | 7.31 | 139 | AFM | 0.00 |
| V2CoRh | -0.239 | 0.058 | 3.76 | 7.20 | 139 | NM | 0.00 |
| Pd2CrHg | -0.042 | 0.058 | 4.11 | 7.81 | 139 | AFM | 0.00 |
| Fe2VMn | -0.086 | 0.058 | 4.05 | 5.73 | 216 | FM | 1.25 |
| Co2NiNb | -0.143 | 0.059 | 3.64 | 7.43 | 119 | FM | 0.42 |
| Co2MnNb | -0.080 | 0.059 | 4.21 | 5.96 | 225 | FM | 1.45 |
| Fe2VTa | -0.124 | 0.059 | 4.21 | 5.96 | 225 | FM | 0.49 |
| Mn2CoV | -0.096 | 0.059 | 4.08 | 5.77 | 225 | FM | 1.09 |
| Ni2MnCo | -0.012 | 0.060 | 3.55 | 7.20 | 139 | FM | 1.45 |
| Co2NiTa | -0.224 | 0.060 | 3.64 | 7.43 | 119 | FM | 0.42 |
| Co2CrPt | -0.031 | 0.060 | 3.62 | 7.37 | 119 | FM | 0.64 |
| Ni2FeNb | -0.171 | 0.060 | 3.64 | 7.56 | 139 | FM | 0.49 |
| Ir2MnZr | -0.537 | 0.060 | 4.28 | 7.05 | 139 | AFM | 0.00 |
| Ir2MnTa | -0.461 | 0.060 | 4.06 | 7.33 | 139 | FM | 0.89 |

| | | | | | | | |
|---|---|---|---|---|---|---|---|
| Ni2MnAu | -0.043 | 0.061 | 3.81 | 7.42 | 119 | AFM | 0.00 |
| Re2FeTi | -0.203 | 0.061 | 4.32 | 6.11 | 216 | FM | 0.51 |
| Rh2MnNb | -0.318 | 0.061 | 4.10 | 7.14 | 139 | FM | 1.02 |
| Ir2FePt | -0.056 | 0.062 | 3.85 | 7.67 | 119 | FM | 0.82 |
| Os2MnNb | -0.033 | 0.062 | 4.40 | 6.22 | 225 | FM | 0.95 |
| Mn2AuRh | -0.063 | 0.062 | 3.98 | 7.48 | 139 | AFM | 0.00 |
| Co2IrTi | -0.365 | 0.062 | 3.69 | 7.33 | 139 | NM | 0.00 |
| Pd2FeAg | -0.013 | 0.063 | 4.03 | 7.50 | 139 | FM | 0.93 |
| Ru2MnTi | -0.306 | 0.063 | 4.30 | 6.08 | 225 | FM | 0.80 |
| Ir2FeNb | -0.366 | 0.063 | 3.94 | 7.58 | 139 | FM | 0.63 |
| Rh2MnTa | -0.400 | 0.063 | 4.14 | 7.04 | 139 | FM | 1.02 |
| Pt2CrAu | -0.106 | 0.064 | 3.99 | 7.87 | 139 | FM | 0.67 |
| V2CoIr | -0.341 | 0.064 | 3.78 | 7.21 | 139 | NM | 0.00 |
| Mn2MoIr | -0.099 | 0.064 | 4.28 | 6.05 | 225 | FM | 1.25 |
| Pt2FeAu | -0.093 | 0.064 | 3.99 | 7.79 | 139 | FM | 0.96 |
| Co2NiTi | -0.270 | 0.064 | 3.62 | 7.21 | 119 | FM | 0.56 |
| Pd2FeZn | -0.274 | 0.064 | 4.01 | 7.12 | 139 | FM | 0.90 |
| Ir2VCo | -0.336 | 0.064 | 3.81 | 7.31 | 139 | FM | 0.30 |
| Ir2MnCu | -0.047 | 0.065 | 3.85 | 7.17 | 139 | FM | 0.53 |
| Fe2TiPt | -0.399 | 0.065 | 3.75 | 7.78 | 119 | FM | 1.09 |
| Fe2VCo | -0.083 | 0.065 | 4.04 | 5.72 | 216 | FM | 1.38 |
| Zn2FeIr | -0.089 | 0.066 | 3.92 | 7.10 | 119 | FM | 0.74 |
| Rh2CoMn | -0.016 | 0.066 | 3.73 | 7.38 | 139 | AFM | 0.00 |
| Ir2FeTa | -0.451 | 0.066 | 3.95 | 7.54 | 139 | FM | 0.65 |
| Fe2CrIr | -0.021 | 0.066 | 3.67 | 7.44 | 119 | FM | 0.73 |
| Ir2CoMo | -0.180 | 0.067 | 3.84 | 7.54 | 139 | FM | 0.35 |
| Ru2CrTa | -0.104 | 0.067 | 4.38 | 6.20 | 225 | FM | 0.72 |
| Pt2CrTi | -0.620 | 0.067 | 3.95 | 7.70 | 139 | AFM | 0.00 |
| Mn2VOs | -0.078 | 0.067 | 4.17 | 5.89 | 216 | FM | 0.72 |
| V2CoOs | -0.191 | 0.067 | 4.20 | 5.94 | 225 | FM | 0.61 |
| Mn2PtNb | -0.256 | 0.067 | 4.36 | 6.16 | 225 | FM | 1.29 |
| Co2VMn | -0.099 | 0.067 | 3.53 | 7.12 | 139 | FERI | 0.24 |
| Pd2MnRh | -0.112 | 0.068 | 3.88 | 7.80 | 139 | AFM | 0.00 |
| Mn2VNi | -0.111 | 0.069 | 4.11 | 5.82 | 225 | FM | 1.21 |
| Co2PtMo | -0.090 | 0.069 | 3.70 | 7.63 | 119 | FM | 0.44 |
| Pt2CrRu | -0.100 | 0.069 | 3.83 | 7.87 | 139 | AFM | 0.00 |
| Co2MoIr | -0.093 | 0.070 | 3.71 | 7.49 | 119 | FM | 0.47 |
| Co2IrNb | -0.231 | 0.070 | 3.75 | 7.63 | 119 | FM | 0.51 |
| Co2TaMn | -0.148 | 0.070 | 4.21 | 5.96 | 225 | FM | 1.44 |
| Rh2CoW | -0.141 | 0.070 | 3.82 | 7.53 | 139 | FM | 0.41 |
| Zn2MnPd | -0.246 | 0.070 | 4.00 | 7.16 | 139 | FM | 0.86 |
| Pd2MnTi | -0.426 | 0.071 | 3.96 | 7.65 | 139 | AFM | 0.00 |
| Rh2CoMo | -0.106 | 0.071 | 3.81 | 7.53 | 139 | FM | 0.41 |
| Fe2RhTi | -0.323 | 0.071 | 4.22 | 5.97 | 216 | FM | 1.23 |
| Zn2MnTi | -0.131 | 0.071 | 3.75 | 8.06 | 139 | FM | 0.32 |
| Rh2FeNb | -0.306 | 0.072 | 3.94 | 7.46 | 139 | FM | 0.71 |
| Pt2CrTa | -0.492 | 0.072 | 3.90 | 8.03 | 139 | AFM | 0.00 |
| Fe2PtMo | -0.087 | 0.072 | 3.74 | 7.69 | 119 | FM | 0.83 |
| Pt2FeZr | -0.730 | 0.072 | 3.98 | 8.04 | 139 | FM | 0.64 |
| Pt2CoZr | -0.729 | 0.072 | 3.94 | 8.07 | 139 | FM | 0.38 |
| Co2TiV | -0.214 | 0.073 | 4.16 | 5.88 | 225 | FM | 0.71 |
| Mn2RuV | -0.071 | 0.074 | 4.16 | 5.89 | 216 | FM | 0.71 |
| Rh2FeTa | -0.384 | 0.074 | 3.96 | 7.42 | 139 | FM | 0.70 |

| | | | | | | | |
|---|---|---|---|---|---|---|---|
| Ag2MnPd | -0.022 | 0.074 | 4.08 | 7.86 | 119 | AFM | 0.00 |
| Ir2MnSc | -0.520 | 0.075 | 4.44 | 6.28 | 225 | FM | 1.01 |
| Co2RhV | -0.175 | 0.075 | 3.65 | 7.29 | 119 | FM | 0.53 |
| Mn2PdIr | -0.110 | 0.075 | 4.36 | 6.16 | 225 | FM | 2.12 |
| Ni2MnNb | -0.158 | 0.076 | 3.67 | 7.61 | 139 | FM | 0.61 |
| V2CoMn | -0.163 | 0.076 | 4.08 | 5.77 | 225 | FM | 0.43 |
| Ti2VCo | -0.138 | 0.076 | 4.30 | 6.09 | 225 | FM | 0.39 |
| Co2IrTa | -0.317 | 0.076 | 3.76 | 7.61 | 119 | FM | 0.50 |
| Cd2MnPt | -0.093 | 0.076 | 4.14 | 8.28 | 139 | FM | 0.99 |
| Rh2CoV | -0.247 | 0.076 | 3.80 | 7.25 | 139 | FM | 0.38 |
| Mn2WRh | -0.015 | 0.076 | 3.71 | 7.77 | 119 | FERRI | 0.17 |
| Pt2CrNb | -0.444 | 0.077 | 3.90 | 8.04 | 139 | AFM | 0.00 |
| Ta2CoRu | -0.201 | 0.077 | 4.17 | 6.99 | 139 | NM | 0.18 |
| Ti2NiZn | -0.216 | 0.078 | 4.37 | 6.18 | 225 | FM | 0.40 |
| V2CoRu | -0.146 | 0.078 | 3.94 | 6.62 | 139 | FM | 0.31 |
| Pd2MnSc | -0.566 | 0.078 | 4.55 | 6.43 | 225 | FM | 1.02 |
| Co2NbSc | -0.181 | 0.079 | 4.38 | 6.20 | 225 | FM | 0.47 |
| Pt2FeIr | -0.078 | 0.080 | 3.90 | 7.66 | 119 | FM | 0.77 |
| Nb2CoRh | -0.200 | 0.080 | 3.99 | 7.55 | 139 | NM | 0.00 |
| Co2RhMo | -0.031 | 0.080 | 3.69 | 7.46 | 119 | FM | 0.48 |
| V2MnIr | -0.332 | 0.080 | 4.24 | 6.00 | 216 | FM | 0.52 |
| V2FeCo | -0.094 | 0.081 | 3.82 | 6.47 | 139 | FERI | 0.19 |
| Fe2MnTa | -0.083 | 0.081 | 4.18 | 5.91 | 225 | FM | 1.12 |
| Pd2FeNb | -0.238 | 0.081 | 3.86 | 8.01 | 139 | FM | 0.54 |
| Mn2VFe | -0.063 | 0.081 | 4.03 | 5.70 | 216 | FM | 0.75 |
| Zr2MnRe | -0.132 | 0.081 | 4.63 | 6.55 | 225 | FM | 0.44 |
| Cd2MnRh | 0.000 | 0.082 | 4.08 | 8.31 | 139 | FM | 1.04 |
| Os2FeTa | -0.065 | 0.082 | 4.39 | 6.21 | 225 | FM | 1.11 |
| Nb2CoRu | -0.082 | 0.082 | 4.17 | 6.98 | 139 | NM | 0.18 |
| Nb2MnCo | -0.050 | 0.083 | 4.34 | 6.14 | 225 | FM | 0.42 |
| Fe2AuPt | -0.034 | 0.084 | 3.88 | 7.73 | 139 | FM | 1.61 |
| Pt2CrIr | -0.160 | 0.084 | 3.87 | 7.78 | 119 | AFM | 0.00 |
| Fe2NbV | -0.051 | 0.084 | 4.22 | 5.97 | 225 | FM | 0.53 |
| Pd2MnY | -0.514 | 0.085 | 4.69 | 6.72 | 139 | FM | 1.00 |
| Ni2FeTi | -0.313 | 0.085 | 3.61 | 7.42 | 139 | FM | 0.61 |
| Ti2CrRu | -0.329 | 0.085 | 4.34 | 6.14 | 225 | FM | 0.37 |
| Ru2MnHf | -0.318 | 0.085 | 4.43 | 6.26 | 225 | FM | 0.85 |
| V2FePt | -0.291 | 0.086 | 3.76 | 7.47 | 139 | FERI | 0.13 |
| Ta2MnCo | -0.172 | 0.086 | 4.34 | 6.14 | 225 | FM | 0.40 |
| Co2FeTa | -0.129 | 0.086 | 3.73 | 7.19 | 139 | FM | 0.84 |
| Ir2CoNb | -0.360 | 0.086 | 3.89 | 7.63 | 139 | FM | 0.37 |
| Co2FeNb | -0.054 | 0.086 | 3.72 | 7.21 | 139 | FM | 0.83 |
| Nb2CoIr | -0.273 | 0.086 | 4.00 | 7.58 | 139 | NM | 0.00 |
| Ni2MnTa | -0.215 | 0.087 | 3.67 | 7.60 | 139 | FM | 0.60 |
| Ir2CrTi | -0.488 | 0.087 | 4.00 | 7.21 | 139 | FM | 0.56 |
| Co2AuTi | -0.206 | 0.088 | 3.73 | 7.79 | 119 | FM | 0.47 |
| Pt2MnAg | -0.173 | 0.088 | 4.02 | 7.79 | 139 | FM | 1.03 |
| Mn2FeTi | -0.125 | 0.088 | 4.10 | 5.80 | 216 | FM | 0.53 |
| Hf2CrOs | -0.320 | 0.088 | 4.59 | 6.50 | 225 | FM | 0.43 |
| Rh2CrV | -0.207 | 0.088 | 3.81 | 7.35 | 139 | FM | 0.29 |
| Ti2CrMn | -0.081 | 0.088 | 4.28 | 6.06 | 225 | FM | 0.60 |
| Fe2NbOs | -0.015 | 0.089 | 4.31 | 6.09 | 216 | FM | 1.28 |
| Fe2VPd | -0.084 | 0.089 | 3.69 | 7.64 | 119 | FM | 0.89 |

| | | | | | | | |
|---|---|---|---|---|---|---|---|
| Pd2FeHg | -0.021 | 0.089 | 4.08 | 7.70 | 139 | FM | 0.91 |
| Pd2FeTi | -0.410 | 0.089 | 3.87 | 7.74 | 139 | FM | 0.60 |
| Fe2NiTa | -0.138 | 0.090 | 3.66 | 7.60 | 119 | FM | 0.93 |
| Pd2FeTa | -0.283 | 0.090 | 3.85 | 8.02 | 139 | FM | 0.53 |
| Mn2CoTi | -0.151 | 0.090 | 4.11 | 5.81 | 216 | FM | 0.75 |
| Mn2TiZr | -0.118 | 0.090 | 4.38 | 6.19 | 225 | FM | 0.39 |
| Au2CrZn | -0.040 | 0.091 | 4.43 | 6.69 | 139 | AFM | 0.00 |
| Ta2FeRh | -0.212 | 0.091 | 4.12 | 7.16 | 139 | NM | 0.01 |
| Pd2MnTa | -0.287 | 0.091 | 3.89 | 8.03 | 139 | FM | 0.68 |
| Fe2WRh | -0.012 | 0.092 | 3.72 | 7.61 | 119 | FM | 0.90 |
| Fe2MnPt | -0.078 | 0.092 | 3.72 | 7.21 | 119 | AFM | 0.00 |
| Ru2VCr | -0.031 | 0.092 | 4.24 | 6.00 | 225 | FM | 0.69 |
| Ir2FeTi | -0.483 | 0.093 | 4.02 | 7.08 | 139 | FM | 0.86 |
| Co2HfPt | -0.485 | 0.093 | 3.84 | 7.79 | 119 | FM | 0.59 |
| Co2NbRh | -0.181 | 0.093 | 3.75 | 7.58 | 119 | FM | 0.50 |
| Mn2VZn | -0.059 | 0.093 | 4.15 | 5.87 | 225 | FM | 0.77 |
| Zr2CrOs | -0.207 | 0.093 | 4.63 | 6.55 | 225 | FM | 0.42 |
| Fe2RhV | -0.128 | 0.094 | 3.68 | 7.50 | 119 | FM | 0.97 |
| Rh2CoNb | -0.301 | 0.094 | 3.88 | 7.53 | 139 | FM | 0.41 |
| Mn2TaOs | -0.077 | 0.094 | 4.32 | 6.10 | 216 | FM | 0.74 |
| Nb2CoOs | -0.094 | 0.094 | 4.25 | 6.83 | 139 | FM | 0.25 |
| Zn2CrPt | -0.244 | 0.094 | 3.98 | 7.36 | 139 | FM | 0.88 |
| Co2ScTa | -0.218 | 0.094 | 4.37 | 6.18 | 225 | FM | 0.41 |
| Ir2CoTa | -0.451 | 0.095 | 3.90 | 7.60 | 139 | FM | 0.39 |
| Zn2MnSc | -0.163 | 0.095 | 4.52 | 6.39 | 225 | FM | 0.83 |
| Ir2CoTi | -0.480 | 0.095 | 3.77 | 7.68 | 119 | FM | 0.36 |
| Rh2FeTi | -0.445 | 0.095 | 4.12 | 6.69 | 139 | FM | 0.95 |
| Pd2CrCd | -0.134 | 0.096 | 4.16 | 7.54 | 139 | AFM | 0.00 |
| Pt2CrAg | -0.074 | 0.096 | 3.97 | 7.87 | 139 | AFM | 0.00 |
| Pd2FeCu | -0.116 | 0.097 | 3.95 | 7.10 | 139 | FM | 0.94 |
| Fe2ZnRh | -0.084 | 0.097 | 4.19 | 5.92 | 216 | FM | 1.45 |
| Pt2MnHg | -0.169 | 0.097 | 4.13 | 7.73 | 139 | AFM | 0.00 |
| Fe2CoTa | -0.091 | 0.097 | 4.19 | 5.92 | 216 | FM | 1.40 |
| Fe2NiNb | -0.070 | 0.097 | 3.66 | 7.62 | 119 | FM | 0.92 |
| Zn2FeTi | -0.125 | 0.098 | 3.69 | 8.06 | 139 | FERI | 0.19 |
| Fe2TaIr | -0.239 | 0.099 | 3.78 | 7.79 | 119 | FM | 1.02 |
| Mn2PtTa | -0.279 | 0.099 | 3.78 | 8.08 | 119 | AFM | 0.00 |
| Ir2CrNb | -0.325 | 0.099 | 3.95 | 7.63 | 139 | AFM | 0.00 |
| Fe2NbIr | -0.167 | 0.099 | 3.84 | 7.43 | 139 | AFM | 0.00 |
| Fe2MnNb | -0.007 | 0.099 | 4.18 | 5.92 | 225 | FM | 1.26 |
| Pt2FeAg | -0.058 | 0.099 | 3.99 | 7.69 | 139 | FM | 0.95 |
| Co2CrTi | -0.133 | 0.099 | 4.13 | 5.85 | 225 | FM | 1.02 |
| Zn2CrIr | -0.072 | 0.100 | 3.91 | 7.34 | 119 | AFM | 0.00 |
| Ru2MnZr | -0.222 | 0.100 | 4.45 | 6.30 | 225 | FM | 0.88 |
| Mn2TiIr | -0.328 | 0.100 | 4.31 | 6.10 | 216 | FM | 0.74 |
| Ta2FeIr | -0.288 | 0.100 | 4.13 | 7.19 | 139 | NM | 0.11 |
| Mn2IrNb | -0.170 | 0.100 | 3.82 | 7.84 | 139 | AFM | 0.00 |
| Zn2FePt | -0.238 | 0.101 | 3.91 | 7.30 | 139 | AFM | 0.00 |
| Mo2MnIr | -0.062 | 0.101 | 3.83 | 7.93 | 139 | FM | 0.47 |
| Co2MnSc | -0.159 | 0.101 | 4.22 | 5.97 | 225 | FM | 1.11 |
| Fe2MnTi | -0.116 | 0.101 | 4.10 | 5.80 | 225 | FM | 1.07 |
| Re2MnV | -0.077 | 0.101 | 4.29 | 6.07 | 216 | FM | 0.46 |
| Ti2FeCr | -0.157 | 0.101 | 4.25 | 6.01 | 225 | FM | 0.40 |

| | | | | | | | |
|---|---|---|---|---|---|---|---|
| Fe2CoNb | -0.025 | 0.101 | 4.19 | 5.92 | 216 | FM | 1.41 |
| Pt2FeCd | -0.169 | 0.102 | 4.07 | 7.68 | 139 | FM | 0.94 |
| Ta2FeCo | -0.134 | 0.102 | 4.34 | 6.14 | 225 | FM | 0.65 |
| Rh2MnY | -0.449 | 0.102 | 4.59 | 6.49 | 225 | FM | 1.10 |
| Rh2CoTa | -0.385 | 0.102 | 3.89 | 7.50 | 139 | FM | 0.42 |
| Co2RhTa | -0.262 | 0.102 | 3.75 | 7.56 | 119 | FM | 0.55 |
| Fe2CrV | -0.012 | 0.103 | 4.04 | 5.71 | 216 | FM | 0.77 |
| Pd2FeCd | -0.132 | 0.104 | 4.11 | 7.47 | 139 | FM | 0.90 |
| Au2MnTi | -0.226 | 0.104 | 4.05 | 8.01 | 139 | FM | 0.64 |
| Pd2FeMn | -0.076 | 0.104 | 3.97 | 7.35 | 139 | FM | 1.84 |
| Mn2VRe | -0.074 | 0.104 | 4.16 | 5.89 | 216 | FM | 0.51 |
| Fe2ZnTi | -0.121 | 0.104 | 4.14 | 5.86 | 225 | FM | 0.51 |
| Fe2MoIr | -0.058 | 0.104 | 3.73 | 7.62 | 119 | FM | 0.90 |
| Ir2CrTa | -0.395 | 0.105 | 3.95 | 7.60 | 139 | FM | 0.40 |
| Zn2FeRh | -0.142 | 0.107 | 3.85 | 7.29 | 119 | FM | 0.74 |
| Ni2CrZn | -0.028 | 0.107 | 3.77 | 6.79 | 139 | FM | 0.72 |
| Mn2HfZr | -0.093 | 0.107 | 4.50 | 6.37 | 225 | FM | 0.40 |
| Co2VTa | -0.122 | 0.107 | 3.82 | 7.11 | 139 | NM | 0.00 |
| Ti2VMn | -0.037 | 0.107 | 4.33 | 6.13 | 225 | FM | 0.51 |
| Ti2FeV | -0.105 | 0.108 | 4.31 | 6.10 | 225 | FM | 0.56 |
| Ti2CrRe | -0.169 | 0.108 | 4.39 | 6.21 | 225 | NM | 0.00 |
| Ru2MnPt | -0.061 | 0.108 | 3.82 | 7.66 | 119 | AFM | 0.00 |
| Pt2FeRu | -0.048 | 0.109 | 3.83 | 7.82 | 139 | FM | 0.90 |
| W2MnIr | -0.044 | 0.110 | 3.86 | 7.88 | 139 | FM | 0.41 |
| Pt2FeSc | -0.713 | 0.110 | 4.04 | 7.68 | 139 | FM | 0.77 |
| Fe2TiW | -0.112 | 0.110 | 4.22 | 5.97 | 225 | FM | 0.49 |
| Fe2NiZn | -0.017 | 0.111 | 4.07 | 5.75 | 216 | FM | 1.29 |
| Co2VNb | -0.066 | 0.111 | 3.81 | 7.13 | 139 | NM | 0.00 |
| Cu2MnTi | -0.044 | 0.112 | 3.63 | 7.95 | 139 | FM | 0.46 |
| Zn2NiFe | -0.041 | 0.113 | 3.78 | 7.07 | 139 | AFM | 0.00 |
| Zn2CrAu | -0.022 | 0.113 | 4.10 | 7.30 | 139 | FM | 0.83 |
| Mn2TaRu | -0.092 | 0.114 | 4.30 | 6.08 | 216 | FM | 0.74 |
| Pd2CrV | -0.097 | 0.114 | 3.85 | 7.75 | 139 | AFM | 0.00 |
| Fe2ZnPt | -0.164 | 0.114 | 3.85 | 7.22 | 119 | FM | 1.36 |
| Pd2CoTi | -0.402 | 0.114 | 3.84 | 7.70 | 139 | FM | 0.35 |
| Os2CrTa | -0.045 | 0.115 | 4.41 | 6.23 | 225 | FM | 0.69 |
| Ru2FeTi | -0.253 | 0.115 | 4.29 | 6.06 | 225 | FM | 1.01 |
| Mn2NbRu | -0.005 | 0.116 | 4.30 | 6.08 | 216 | FM | 0.73 |
| Co2RhTi | -0.295 | 0.116 | 3.65 | 7.71 | 139 | FM | 0.74 |
| Fe2AuTi | -0.158 | 0.116 | 3.77 | 7.95 | 119 | FM | 0.97 |
| Co2TiPd | -0.275 | 0.116 | 3.74 | 7.46 | 119 | FM | 0.58 |
| Nb2MnIr | -0.233 | 0.117 | 4.46 | 6.30 | 216 | FM | 0.57 |
| Hf2MnMo | -0.048 | 0.117 | 4.63 | 6.55 | 225 | FM | 0.58 |
| Co2FeSc | -0.148 | 0.117 | 4.22 | 5.97 | 225 | FM | 1.30 |
| Nb2MnPt | -0.267 | 0.117 | 3.92 | 8.37 | 139 | FM | 0.69 |
| Ni2MnTi | -0.277 | 0.117 | 3.69 | 7.29 | 139 | FM | 0.80 |
| Pt2MnY | -0.705 | 0.118 | 4.90 | 6.12 | 139 | FM | 1.08 |
| Fe2CoHf | -0.166 | 0.118 | 4.24 | 6.00 | 216 | FM | 1.30 |
| Pt2MnRu | -0.143 | 0.118 | 3.84 | 7.85 | 139 | FM | 0.69 |
| Fe2NiTi | -0.189 | 0.119 | 3.55 | 7.91 | 139 | FM | 1.16 |
| Zn2FeSc | -0.156 | 0.119 | 3.89 | 7.95 | 139 | FM | 0.31 |
| Co2NbTa | -0.085 | 0.119 | 4.01 | 7.09 | 139 | AFM | 0.00 |
| Co2FeHf | -0.173 | 0.119 | 4.26 | 6.02 | 225 | FM | 1.33 |

| | | | | | | | |
|---|---|---|---|---|---|---|---|
| Pt2CrSc | -0.703 | 0.120 | 4.17 | 7.52 | 139 | AFM | 0.00 |
| Co2TaPd | -0.201 | 0.120 | 3.77 | 7.91 | 119 | FM | 0.31 |
| Os2MnTi | -0.236 | 0.121 | 4.31 | 6.10 | 225 | FM | 0.74 |
| Co2TiRe | -0.144 | 0.121 | 3.74 | 7.23 | 139 | NM | 0.00 |
| Zn2MnHf | -0.082 | 0.121 | 3.92 | 8.05 | 139 | FM | 0.41 |
| Co2AuHf | -0.250 | 0.121 | 3.81 | 8.10 | 119 | FM | 0.45 |
| Ni2TiV | -0.226 | 0.121 | 3.70 | 7.39 | 139 | NM | 0.00 |
| Ir2MnPd | -0.050 | 0.121 | 3.86 | 7.65 | 139 | AFM | 0.00 |
| Re2MnTa | -0.045 | 0.122 | 4.38 | 6.19 | 216 | FM | 0.53 |
| Re2CoTi | -0.150 | 0.122 | 4.31 | 6.10 | 216 | FM | 0.41 |
| Ni2MnSc | -0.269 | 0.122 | 4.30 | 6.09 | 225 | FM | 1.06 |
| V2FeRh | -0.161 | 0.122 | 3.86 | 6.89 | 139 | NM | 0.02 |
| Cd2CrPt | -0.030 | 0.122 | 4.17 | 8.22 | 139 | FM | 0.93 |
| V2MnRh | -0.217 | 0.122 | 4.24 | 5.99 | 216 | FM | 0.53 |
| Rh2CrSc | -0.422 | 0.122 | 4.50 | 6.10 | 139 | AFM | 0.00 |
| Cr2MnPt | -0.047 | 0.123 | 3.67 | 7.56 | 139 | FERI | 0.11 |
| Cu2CoPt | -0.001 | 0.123 | 3.74 | 7.32 | 119 | FM | 0.53 |
| Ag2MnPt | -0.047 | 0.123 | 4.11 | 7.74 | 119 | AFM | 0.00 |
| Pd2MnZr | -0.479 | 0.123 | 3.95 | 8.42 | 119 | FM | 0.89 |
| Zn2MnPt | -0.327 | 0.123 | 3.96 | 7.34 | 139 | FM | 0.89 |
| Fe2MoTi | -0.093 | 0.124 | 4.22 | 5.97 | 225 | FM | 0.53 |
| Pt2CoSc | -0.699 | 0.124 | 3.98 | 7.77 | 139 | FM | 0.47 |
| Co2MnPt | -0.045 | 0.124 | 3.71 | 7.40 | 119 | FM | 1.67 |
| Ta2CrPt | -0.204 | 0.125 | 3.98 | 8.20 | 139 | FM | 0.50 |
| Zn2MnZr | -0.114 | 0.125 | 4.51 | 6.38 | 225 | FM | 0.75 |
| Nb2FePt | -0.257 | 0.125 | 3.95 | 7.97 | 139 | AFM | 0.00 |
| Ta2MnPt | -0.290 | 0.125 | 3.95 | 8.17 | 139 | FM | 0.50 |
| Fe2ReTi | -0.116 | 0.125 | 4.19 | 5.96 | 225 | FM | 0.73 |
| Hg2MnSc | -0.095 | 0.126 | 4.84 | 6.85 | 225 | FM | 0.97 |
| Sc2MnIr | -0.475 | 0.126 | 4.67 | 6.34 | 139 | FM | 0.68 |
| Mn2TaIr | -0.220 | 0.126 | 4.33 | 6.12 | 216 | FM | 0.97 |
| Rh2CoTi | -0.414 | 0.126 | 3.89 | 7.25 | 139 | FM | 0.48 |
| V2FeIr | -0.264 | 0.127 | 3.80 | 7.23 | 139 | FERI | 0.25 |
| Fe2RhNb | -0.113 | 0.127 | 3.76 | 7.82 | 119 | FM | 1.07 |
| Ir2CrHf | -0.523 | 0.127 | 4.14 | 7.35 | 139 | AFM | 0.00 |
| Au2MnNb | -0.011 | 0.127 | 4.08 | 8.06 | 139 | FM | 0.61 |
| Zr2NiZn | -0.252 | 0.128 | 4.65 | 6.57 | 225 | FM | 0.30 |
| Pt2CrHf | -0.712 | 0.129 | 3.95 | 8.38 | 119 | AFM | 0.00 |
| Cd2MnPd | -0.072 | 0.129 | 4.20 | 8.09 | 139 | FM | 0.98 |
| Fe2HfIr | -0.351 | 0.129 | 4.37 | 6.19 | 216 | FM | 1.29 |
| Co2IrHf | -0.350 | 0.130 | 3.82 | 7.56 | 139 | NM | 0.00 |
| Nb2MnRh | -0.139 | 0.130 | 4.45 | 6.30 | 216 | FM | 0.58 |
| Fe2NbPd | -0.081 | 0.130 | 3.76 | 7.94 | 119 | FM | 0.96 |
| Fe2TaRh | -0.177 | 0.130 | 3.76 | 7.80 | 119 | FM | 1.06 |
| Co2PtZr | -0.422 | 0.131 | 3.85 | 7.87 | 119 | FM | 0.59 |
| Os2VCr | -0.020 | 0.131 | 4.26 | 6.03 | 225 | FM | 0.63 |
| Fe2PdTa | -0.134 | 0.131 | 3.76 | 7.92 | 119 | FM | 0.97 |
| Zn2CrRh | -0.117 | 0.131 | 3.88 | 7.53 | 139 | FM | 0.87 |
| Y2MnZn | -0.054 | 0.131 | 5.04 | 7.13 | 225 | FM | 0.79 |
| Hf2CrRe | -0.144 | 0.132 | 4.64 | 6.56 | 225 | FM | 0.63 |
| Pt2CrHg | -0.044 | 0.132 | 4.06 | 7.95 | 139 | AFM | 0.00 |
| Mn2TiRh | -0.258 | 0.133 | 4.26 | 6.02 | 216 | FM | 0.72 |
| Co2CuHf | -0.149 | 0.133 | 3.69 | 7.83 | 119 | FM | 0.40 |

| | | | | | | | |
|---|---|---|---|---|---|---|---|
| Rh2FeHf | -0.516 | 0.134 | 4.22 | 6.90 | 139 | FM | 0.94 |
| Pd2MnIr | -0.060 | 0.134 | 3.89 | 7.69 | 119 | AFM | 0.00 |
| Pt2CrZr | -0.668 | 0.134 | 3.97 | 8.40 | 119 | AFM | 0.00 |
| Zr2CrRu | -0.203 | 0.134 | 4.63 | 6.54 | 225 | FM | 0.41 |
| Mo2CrPt | -0.024 | 0.135 | 3.84 | 8.08 | 139 | FM | 0.46 |
| Fe2TiCu | -0.081 | 0.136 | 3.63 | 7.60 | 119 | FM | 0.90 |
| Ir2FeSc | -0.425 | 0.136 | 4.42 | 6.25 | 225 | FM | 1.14 |
| Mn2IrZn | -0.035 | 0.136 | 3.79 | 7.57 | 119 | AFM | 0.00 |
| Hf2MnCr | -0.007 | 0.137 | 4.54 | 6.41 | 225 | FM | 0.65 |
| Cu2MnSc | -0.055 | 0.137 | 4.42 | 6.25 | 225 | FM | 0.93 |
| Co2WTi | -0.096 | 0.137 | 3.90 | 6.92 | 139 | AFM | 0.00 |
| Mn2PtAg | -0.033 | 0.137 | 3.95 | 7.83 | 225 | FM | 2.03 |
| Mn2VPd | -0.104 | 0.137 | 4.29 | 6.07 | 225 | FM | 1.23 |
| Zn2FeHf | -0.098 | 0.137 | 3.79 | 8.31 | 139 | NM | 0.07 |
| Pd2MnMo | -0.043 | 0.137 | 3.85 | 8.00 | 139 | AFM | 0.00 |
| Pt2CoCd | -0.059 | 0.137 | 3.99 | 7.81 | 139 | FM | 0.58 |
| Hf2CrRu | -0.309 | 0.138 | 4.58 | 6.48 | 225 | FM | 0.42 |
| Mn2FeTa | -0.033 | 0.138 | 4.18 | 5.92 | 216 | FM | 0.77 |
| Ni2MnZr | -0.231 | 0.139 | 3.75 | 8.01 | 119 | FM | 0.92 |
| Fe2PtRe | -0.019 | 0.139 | 3.72 | 7.61 | 119 | FM | 0.84 |
| Sc2CrRu | -0.168 | 0.140 | 4.62 | 6.54 | 225 | FM | 0.82 |
| Cu2MnHf | -0.036 | 0.140 | 3.79 | 8.00 | 139 | FM | 0.55 |
| Rh2FeZr | -0.436 | 0.141 | 4.22 | 7.01 | 139 | FM | 0.94 |
| Sc2CrOs | -0.136 | 0.142 | 4.62 | 6.53 | 225 | FM | 0.81 |
| Au2MnHf | -0.297 | 0.142 | 4.68 | 6.61 | 225 | FM | 0.99 |
| Ru2FeHf | -0.261 | 0.142 | 4.41 | 6.24 | 225 | FM | 1.03 |
| Au2MnZr | -0.310 | 0.143 | 4.71 | 6.66 | 225 | FM | 1.01 |
| Fe2CoZr | -0.098 | 0.143 | 4.27 | 6.03 | 216 | FM | 1.32 |
| Co2PdNb | -0.131 | 0.143 | 3.74 | 7.72 | 119 | FM | 0.47 |
| Mn2CoTa | -0.051 | 0.144 | 4.19 | 5.93 | 216 | FM | 1.01 |
| V2MnPt | -0.281 | 0.144 | 3.72 | 7.89 | 139 | FM | 0.54 |
| Co2VHf | -0.177 | 0.145 | 4.30 | 6.08 | 225 | FM | 0.75 |
| Co2MoTi | -0.088 | 0.145 | 3.82 | 7.12 | 139 | NM | 0.00 |
| Co2NiZr | -0.191 | 0.145 | 3.77 | 7.47 | 119 | FM | 0.57 |
| Hf2FeCr | -0.073 | 0.145 | 4.50 | 6.37 | 225 | FM | 0.44 |
| Fe2AuTa | -0.012 | 0.145 | 3.79 | 7.99 | 119 | FM | 0.84 |
| Mn2TaZn | -0.026 | 0.145 | 4.26 | 6.02 | 225 | FM | 0.77 |
| Sc2FeOs | -0.192 | 0.145 | 4.53 | 6.41 | 225 | FM | 0.44 |
| Nb2MnAu | -0.013 | 0.145 | 3.95 | 8.50 | 139 | AFM | 0.00 |
| Pd2MnHf | -0.512 | 0.145 | 3.92 | 8.40 | 119 | FM | 0.90 |
| Rh2CrTi | -0.395 | 0.145 | 4.04 | 7.05 | 139 | AFM | 0.00 |
| Fe2TiPd | -0.213 | 0.145 | 3.75 | 7.77 | 119 | FM | 1.08 |
| Co2FeZr | -0.109 | 0.145 | 4.28 | 6.06 | 225 | FM | 1.37 |
| Mn2AuTi | -0.123 | 0.146 | 3.80 | 8.37 | 119 | FM | 1.35 |
| Mn2RhNb | -0.098 | 0.146 | 4.32 | 6.11 | 225 | FM | 1.22 |
| Cu2MnZr | -0.014 | 0.147 | 3.81 | 8.05 | 139 | FM | 0.56 |
| Mn2PdNb | -0.068 | 0.147 | 4.36 | 6.17 | 225 | FM | 1.29 |
| Y2MnCd | -0.035 | 0.148 | 5.23 | 7.21 | 139 | FM | 0.82 |
| Pt2CoPd | -0.044 | 0.148 | 3.87 | 7.79 | 139 | FM | 0.71 |
| Cr2VIr | -0.119 | 0.148 | 4.21 | 5.95 | 216 | FM | 0.42 |
| Fe2TaRe | -0.008 | 0.149 | 4.28 | 6.05 | 225 | FM | 1.00 |
| Ni2FeZr | -0.244 | 0.151 | 3.77 | 7.61 | 139 | FM | 0.62 |
| Ti2MnAu | -0.209 | 0.151 | 3.91 | 8.28 | 119 | FM | 0.34 |

| | | | | | | | |
|---|---|---|---|---|---|---|---|
| Sc2MnZn | -0.090 | 0.151 | 4.60 | 6.96 | 139 | FM | 0.64 |
| Co2FeCu | -0.093 | 0.152 | 3.71 | 7.87 | 119 | FM | 0.39 |
| Zn2FePd | -0.164 | 0.152 | 3.95 | 7.13 | 139 | AFM | 0.00 |
| Mn2NiNb | -0.019 | 0.152 | 4.22 | 5.97 | 225 | FM | 1.26 |
| Ir2CoHf | -0.498 | 0.152 | 3.96 | 7.71 | 139 | FM | 0.48 |
| Rh2CrNb | -0.204 | 0.153 | 3.92 | 7.59 | 139 | FM | 0.46 |
| Au2FeTi | -0.177 | 0.154 | 3.99 | 7.99 | 139 | FM | 0.48 |
| Ru2FeZr | -0.167 | 0.154 | 4.43 | 6.27 | 225 | FM | 1.05 |
| Mn2TaSc | -0.030 | 0.154 | 4.39 | 6.21 | 225 | FM | 0.43 |
| Nb2FeRh | -0.109 | 0.156 | 4.08 | 7.26 | 139 | NM | 0.00 |
| Y2MnAg | -0.017 | 0.156 | 5.11 | 7.23 | 225 | FM | 0.85 |
| Ir2CrZr | -0.441 | 0.157 | 4.15 | 7.42 | 139 | AFM | 0.00 |
| Mn2TiZn | -0.052 | 0.159 | 4.24 | 5.99 | 225 | FM | 0.96 |
| Zn2MnY | -0.120 | 0.159 | 4.71 | 6.66 | 225 | FM | 0.90 |
| Pt2MnMo | -0.308 | 0.159 | 3.87 | 7.96 | 139 | AFM | 0.00 |
| Fe2CrTi | -0.058 | 0.160 | 4.11 | 5.82 | 216 | FM | 0.54 |
| Y2MnHg | -0.101 | 0.160 | 5.25 | 7.08 | 139 | FM | 0.81 |
| Co2AuZr | -0.205 | 0.161 | 3.82 | 8.17 | 119 | FM | 0.46 |
| Sc2CoMn | -0.075 | 0.162 | 4.61 | 6.24 | 139 | FM | 0.82 |
| Fe2RhHf | -0.313 | 0.162 | 4.35 | 6.15 | 216 | FM | 1.23 |
| Sc2CrIr | -0.424 | 0.163 | 4.63 | 6.55 | 225 | FM | 0.75 |
| Mn2TaRe | -0.008 | 0.163 | 4.30 | 6.08 | 216 | FM | 0.63 |
| Au2MnSc | -0.457 | 0.163 | 4.68 | 6.62 | 225 | FM | 1.01 |
| Rh2CoZr | -0.414 | 0.164 | 4.01 | 7.51 | 139 | FM | 0.52 |
| Co2CrHf | -0.118 | 0.164 | 4.27 | 6.04 | 225 | FM | 1.04 |
| Co2TiOs | -0.192 | 0.164 | 3.71 | 7.30 | 139 | NM | 0.00 |
| Rh2CoHf | -0.486 | 0.165 | 4.01 | 7.43 | 139 | FM | 0.51 |
| Co2CrTa | -0.027 | 0.165 | 4.21 | 5.96 | 225 | FM | 1.20 |
| Rh2CrTa | -0.272 | 0.165 | 3.93 | 7.56 | 139 | FM | 0.44 |
| Sc2MnCu | -0.023 | 0.166 | 4.73 | 6.26 | 139 | FM | 0.58 |
| Pd2FeZr | -0.461 | 0.166 | 3.97 | 8.05 | 139 | FM | 0.62 |
| Sc2MnHg | -0.101 | 0.166 | 4.66 | 7.45 | 139 | FM | 0.70 |
| Co2VSc | -0.132 | 0.166 | 4.25 | 6.02 | 225 | FM | 0.52 |
| Ir2CrPd | -0.009 | 0.166 | 3.85 | 7.62 | 139 | NM | 0.23 |
| Cr2VRh | -0.028 | 0.167 | 4.21 | 5.96 | 216 | FM | 0.48 |
| Rh2CrHf | -0.483 | 0.167 | 4.48 | 6.33 | 225 | FM | 0.96 |
| Sc2MnRh | -0.390 | 0.167 | 4.68 | 6.29 | 139 | FM | 0.62 |
| Pt2FeRe | -0.055 | 0.168 | 3.80 | 7.94 | 139 | FM | 0.58 |
| Nb2CrPt | -0.181 | 0.169 | 3.95 | 8.32 | 139 | FM | 0.56 |
| Fe2TiRu | -0.199 | 0.169 | 4.21 | 5.96 | 216 | FM | 1.22 |
| Sc2CrZn | -0.020 | 0.170 | 4.75 | 6.72 | 225 | FM | 0.80 |
| Co2IrZr | -0.278 | 0.170 | 3.86 | 7.65 | 139 | FM | 0.71 |
| Mn2TaRh | -0.146 | 0.171 | 4.31 | 6.10 | 216 | FM | 0.98 |
| Nb2FeIr | -0.176 | 0.171 | 4.03 | 7.58 | 139 | FM | 0.32 |
| Ir2CoZr | -0.425 | 0.172 | 3.96 | 7.79 | 139 | FM | 0.46 |
| Cu2FeSc | -0.027 | 0.172 | 3.77 | 7.94 | 139 | FM | 0.46 |
| Zn2FeZr | -0.092 | 0.172 | 3.97 | 7.81 | 139 | FM | 0.32 |
| Mn2AuZr | -0.127 | 0.172 | 3.90 | 8.64 | 119 | FM | 1.41 |
| Sc2MnPt | -0.555 | 0.173 | 4.56 | 6.82 | 139 | AFM | 0.00 |
| Cd2MnSc | -0.016 | 0.173 | 4.83 | 6.83 | 225 | FM | 0.91 |
| Sc2MnAg | -0.020 | 0.173 | 4.88 | 6.47 | 139 | FM | 0.64 |
| Co2VZr | -0.106 | 0.173 | 4.32 | 6.11 | 225 | FM | 0.76 |
| Pd2FeSc | -0.471 | 0.173 | 4.06 | 7.63 | 139 | FM | 0.73 |

| | | | | | | |
|---|---|---|---|---|---|---|
| Sc2MnCd | -0.026 | 0.174 | 4.77 | 7.19 | 139 | FM | 0.71 |
| Ru2FeSc | -0.105 | 0.174 | 4.37 | 6.17 | 225 | FM | 0.84 |
| Cr2TaIr | -0.127 | 0.175 | 4.36 | 6.17 | 216 | FM | 0.46 |

Table S. 2- All-*d*-metal Heuslers from literature, with the results of the HTP search and the available experimental lattice parameters.

| Compound | Exp. Latt. Param. [Å] | Exp. c/a | Latt. Param. [Å] | c/a | $E_f$ [ev/atom] | $\Delta E_h$ [ev/atom] |
|---|---|---|---|---|---|---|
| Ni2MnTi[a] | 5.93 (RT) | 1.00 (RT) | 5.93 | 1.00 | -0.275 | 0.117 |
| Co2MnV[b] | --- | --- | 3.53 | 1.416 | -0.99 | 0.067 |
| Co2MnTi[c] | 5.89 | 1.00 | 5.83 | 1.000 | -0.275 | 0.00 |
| Mn2PdPt[c] | 4.03 | 1.27 | 4.23 | 1.13 | -0.274 | 0.00 |
| Mn2PtCo[c] | Unstable | --- | 6.01 | 1.00 | -0.160 | 0.010 |
| Mn2PtV[c] | Unstable | --- | 6.08 | 1.00 | -0.292 | 0.028 |

a-10.1063/1.4927058
b-10.1002/anie.202013610
c-10.1126/sciadv.1602241

Table S. 3- List of Heusler with main-group elements and 3d magnetic atoms from literature, with their ground structure regular (R) and inverse (I).

| Composition | Type | DOI | Composition | Type | DOI |
|---|---|---|---|---|---|
| Au2MnAl | R | https://doi.org/10.1107/S0365110X63002164 | Ni2CuSn | R | https://doi.org/10.1016/j.ssc.2015.08.022 |
| Co2CrAl | R | https://doi.org/10.1002/pssb.202100533 | Ni2HfGa | R | https://doi.org/10.1016/j.jallcom.2015.11.126 |
| Co2CrGa | R | https://doi.org/10.1063/1.1790029 | Ni2HfIn | R | https://doi.org/10.1016/j.jallcom.2015.11.126 |
| Co2CrIn | R | https://doi.org/10.1515/znb-2006-0615 | Ni2MgSb | R | https://doi.org/10.1002/zaac.200900036 |
| Co2FeAl | R | https://doi.org/10.1063/1.4993698 | Ni2MnAl | R | https://doi.org/10.1063/1.1504498 |
| Co2FeGa | R | https://doi.org/10.1088/0022-3727/42/8/084018 | Ni2MnGa | R | https://doi.org/10.1016/j.scriptamat.2009.04.046 |
| Co2FeGe | R | https://doi.org/10.1016/j.jallcom.2021.160341 | Ni2MnGe | R | https://doi.org/10.1016/j.jallcom.2014.08.149 |
| Co2FeIn | R | https://doi.org/10.1016/0304-8853(83)90097-5 | Ni2MnIn | R | https://doi.org/10.1016/j.jmmm.2004.11.272 |
| Co2FeSi | R | https://doi.org/10.1063/1.2166205 | Ni2MnSb | R | https://doi.org/10.1021/acsomega.9b03386 |
| Co2HfAl | R | https://doi.org/10.1007/BF01020967 | Ni2MnSn | R | https://doi.org/10.1088/1361-6463/ac4b59 |
| Co2HfGa | R | https://doi.org/10.1007/BF01020967 | Ni2NbAl | R | https://doi.org/10.1143/JPSJ.54.1673 |
| Co2HfSn | R | https://doi.org/10.3390/met10050624 | Ni2NbGa | R | https://doi.org/10.1143/JPSJ.54.1673 |
| Co2MnAl | R | https://doi.org/10.1038/s41467-020-17174-9 | Ni2NbSn | R | https://doi.org/10.1143/JPSJ.54.1673 |
| Co2MnGa | R | https://doi.org/10.1038/s41427-019-0116-z | Ni2ScAl | R | https://doi.org/10.1063/1.353704 |
| Co2MnGe | R | https://doi.org/10.1063/1.373373 | Ni2ScSn | R | https://doi.org/10.1063/1.353704 |
| Co2MnSb | R | https://doi.org/10.1063/1.3054291 | Ni2TaAl | R | https://doi.org/10.1016/j.intermet.2019.106559 |
| Co2MnSi | R | https://doi.org/10.1016/j.jallcom.2018.12.018 | Ni2TiAl | R | https://doi.org/10.1016/j.actamat.2003.08.003 |
| Co2MnSn | R | https://doi.org/10.1088/1742-6596/200/6/062012 | Ni2TiGa | R | https://doi.org/10.1016/j.jallcom.2015.11.126 |
| Co2NbGa | R | https://doi.org/10.1016/j.jmmm.2020.166604 | Ni2TiIn | R | https://doi.org/10.1016/j.jallcom.2015.11.126 |
| Co2NbSn | R | https://doi.org/10.1103/PhysRevB.66.174428 | Ni2TiSn | R | https://doi.org/10.1016/j.jallcom.2015.11.126 |
| Co2ScSn | R | https://doi.org/10.1063/1.361616 | Ni2VAl | R | https://doi.org/10.1016/S0921-4526(99)00102-7 |
| Co2TiAl | R | https://doi.org/10.1016/j.jallcom.2006.05.083 | Ni2VGa | R | https://doi.org/10.1016/j.jmmm.2020.167083 |
| Co2TiGa | R | https://doi.org/10.1088/1742-6596/200/6/062018 | Ni2ZrGa | R | https://doi.org/10.1103/PhysRevB.78.184506 |
| Co2TiGe | R | https://doi.org/10.1038/s41598-019-39037-0 | Ni2ZrSn | R | https://doi.org/10.1002/pssa.2211540207 |
| Co2TiSi | R | http://dx.doi.org/10.1063/1.4984311 | Pd2MnGe | R | https://doi.org/10.1002/pssa.19700030414 |
| Co2TiSn | R | https://doi.org/10.1016/S0304-8853(96)00338-1 | Pd2MnIn | R | https://doi.org/10.1088/1361-648X/aaddd7 |
| Co2VAl | R | https://doi.org/10.1103/PhysRevB.82.144415 | Pd2MnSb | R | https://doi.org/10.1063/1.2163486 |
| Co2VGa | R | https://doi.org/10.3390/ma15176138 | Pd2MnSn | R | https://doi.org/10.1063/1.2163486 |
| Co2VSi | R | https://doi.org/10.1016/j.msea.2016.08.083 | Pt2MnAl | R | https://doi.org/10.1063/1.1660242 |
| Co2VSn | R | https://doi.org/10.1016/S0022-3697(72)80437-2 | Rh2MnAl | R | https://doi.org/10.1143/JPSJ.32.281 |
| Co2ZrAl | R | https://doi.org/10.1016/j.jallcom.2004.09.071 | Rh2MnGe | R | https://doi.org/10.1088/0022-3727/42/8/084001 |
| Co2ZrSn | R | https://doi.org/10.1016/S0304-8853(96)00338-1 | Rh2MnPb | R | https://doi.org/10.1063/1.326881 |
| Cu2CrAl | R | https://doi.org/10.1088/0953-8984/13/23/310 | Rh2MnSn | R | https://doi.org/10.1016/0038-1098(77)91003-1 |

| Composition | | Reference | Composition | | Reference |
|---|---|---|---|---|---|
| Cu2FeSn | R | https://doi.org/10.1002/pssa.200306809 | Ru2FeSi | R | https://doi.org/10.1016/0304-8853(85)90036-8 |
| Cu2MnAl | R | https://doi.org/10.1016/j.jmmm.2006.07.027 | Ru2FeSn | R | https://doi.org/10.1103/PhysRevMaterials.5.064417 |
| Cu2MnIn | R | https://doi.org/10.1002/pssa.19700030414 | Cr2CoGa | I | https://doi.org/10.1002/zaac.200900036 |
| Cu2MnSb | R | https://doi.org/10.1038/194465a0 | Fe2CoGa | I | https://doi.org/10.1039/D2TC02729A |
| Cu2MnSn | R | https://doi.org/10.1016/j.intermet.2017.02.014 | Fe2CoGe | I | https://doi.org/10.1103/PhysRevB.87.064411 |
| Cu2NiSn | R | https://doi.org/10.1002/pssa.2210580254 | Fe2CrGa | I | https://doi.org/10.1063/5.0105273 |
| Cu2FeSn | R | http://jetp.ras.ru/cgi-bin/dn/e_023_02_0219.pdf | Fe2NiAl | I | https://doi.org/10.1016/j.physb.2013.04.005 |
| Fe2CrAl | R | https://doi.org/10.1088/0953-8984/13/23/310 | Fe2NiGa | I | https://doi.org/10.1016/j.physb.2015.08.027 |
| Fe2MnAl | R | https://doi.org/10.1016/j.jmmm.2018.03.018 | Mn2CoAl | I | https://doi.org/10.1103/PhysRevB.77.014424 |
| Fe2MnSi | R | https://doi.org/10.1016/j.jmmm.2018.11.113 | Mn2CoGa | I | https://doi.org/10.1103/PhysRevB.77.014424 |
| Fe2TiAl | R | https://doi.org/10.1016/j.jallcom.2004.01.035 | Mn2CoGe | I | https://doi.org/10.1103/PhysRevB.77.014424 |
| Fe2TiSn | R | https://doi.org/10.1021/acsami.2c04474 | Mn2CoIn | I | https://doi.org/10.1103/PhysRevB.77.014424 |
| Fe2VAl | R | https://doi.org/10.1016/j.actamat.2016.08.080 | Mn2CoSb | I | https://doi.org/10.1103/PhysRevB.77.014424 |
| Fe2VGa | R | https://doi.org/10.1103/PhysRevB.80.235121 | Mn2CoSn | I | https://doi.org/10.1103/PhysRevB.77.014424 |
| Fe2VSi | R | https://doi.org/10.1063/1.5007322 | Mn2FeSi | I | https://doi.org/10.3390/ma15030697 |
| Mn2RhGa | R | https://doi.org/10.1002/zaac.201300665 | Mn2NiIn | I | https://doi.org/10.1002/zaac.201300665 |
| Mn2RuGe | R | https://doi.org/10.1002/zaac.201300665 | Mn2NiSb | I | https://doi.org/10.1002/zaac.201300665 |
| Mn2RuSn | R | https://doi.org/10.1002/zaac.201300665 | Mn2RuSi | I | https://doi.org/10.1002/zaac.201300665 |
| Mn2VAl | R | https://doi.org/10.1002/zaac.201300665 | Ni2CuAl | I | https://doi.org/10.1016/j.jallcom.2015.11.126 |
| Mn2VGa | R | https://doi.org/10.1002/zaac.201300665 | Mn2NiSn | I | https://doi.org/10.1002/zaac.201300665 |
| Ni2CuSb | R | https://doi.org/10.1016/j.ssc.2015.08.022 | Mn2NiGa | I | https://doi.org/10.1002/zaac.201300665 |

Table S. 4- List of all-*d*-metal Heusler that are dynamically unstable from phonon calculations.

| Composition | SPG | State | Composition | State | SPG | Composition | SPG | State |
|---|---|---|---|---|---|---|---|---|
| Cd2MnPd | 139 | FM | Ta2RuCo | 139 | FM | Rh2ScCr | 139 | AFM |
| Cd2MnPt | 139 | FM | Ta2CrPt | 139 | FM | Rh2TiCr | 139 | AFM |
| Cd2MnRh | 139 | FM | Ta2FeIr | 139 | FM | Rh2MnCd | 139 | AFM |
| Cu2HfMn | 139 | FM | Ta2FePt | 139 | FM | Ir2ScMn | 225 | FM |
| Fe2CrRh | 139 | FM | Ta2FeRh | 139 | FM | Mn2HfZr | 225 | FM |
| Fe2RhPd | 139 | FM | Ta2MnPt | 139 | FM | Mn2TiZn | 225 | FM |
| Ir2TiFe | 139 | FM | V2CoIr | 139 | FM | Os2VCr | 225 | FM |
| Mn2CrPt | 139 | FM | V2CoRh | 139 | FM | Ru2VCr | 225 | FM |
| Nb2CoIr | 139 | FM | V2RuCo | 139 | FM | Ti2NiZn | 225 | FM |
| Nb2OsCo | 139 | FM | V2FeCo | 139 | FM | Y2MnZn | 225 | FM |
| Nb2RuCo | 139 | FM | V2FeIr | 139 | FM | Zn2ZrMn | 225 | FM |
| Nb2CrPt | 139 | FM | V2FePt | 139 | FM | Zr2NiZn | 225 | FM |
| Nb2FeIr | 139 | FM | V2FeRh | 139 | FM | Co2HfTi | 225 | FM |
| Nb2FeRh | 139 | FM | Y2MnHg | 139 | FM | Co2HfV | 225 | FM |
| Pt2CrFe | 139 | FM | Zn2HfFe | 139 | FM | Co2ZrV | 225 | FM |
| Pt2YMn | 139 | FM | Co2TiIr | 139 | FM | Hg2YMn | 225 | AFM |
| Rh2TaCr | 139 | FM | Co2ZrIr | 139 | FM | Ti2MoMn | 225 | AFM |
| Rh2HfFe | 139 | FM | Co2TiMo | 139 | FM | Y2MnPd | 225 | AFM |
| Rh2TiFe | 139 | FM | Co2NbV | 139 | FM | Co2CrPt | 119 | FM |
| Rh2ZrFe | 139 | FM | Co2TaV | 139 | FM | Mn2ZrAu | 119 | FM |
| Sc2MnAg | 139 | FM | Ir2VCr | 139 | AFM | Co2MnIr | 119 | AFM |
| Sc2MnCd | 139 | FM | Mn2PtAu | 139 | AFM | Cr2VIr | 216 | FM |
| Sc2MnHg | 139 | FM | Nb2CoRh | 139 | AFM | --- | --- | --- |
| Sc2MnZn | 139 | FM | Nb2FePt | 139 | AFM | --- | --- | --- |

Table S. 5- List of all-*d*-metal Heusler that are mechanically unstable from elastic constants.

| Composition | SPG | State | Composition | State | SPG | Composition | State | SPG |
|---|---|---|---|---|---|---|---|---|
| Fe2CrRh | 139 | FM | Ta2MnPt | 139 | FM | Y2MnPd | 225 | AFM |
| Fe2TiNi | 139 | FM | V2MnPt | | FM | Co2HfNb | 225 | AFM |
| Mn2FePt | 139 | FM | Zn2HfFe | | FM | Co2TiNb | 225 | AFM |
| Mn2CrPt | 139 | FM | Zn2ZrFe | 139 | FM | Co2HfTa | 225 | AFM |

| | | | | | | | | |
|---|---|---|---|---|---|---|---|---|
| Nb2CrPt | 139 | FM | Mn2PdAu | 225 | FM | Co2TiTa | 225 | AFM |
| Nb2FeRh | 139 | FM | Rh2HfCr | 225 | FM | Co2ZrTa | 225 | AFM |
| Nb2MnAu | 139 | FM | Co2HfV | 225 | FM | Fe2TiAu | 119 | FM |
| Nb2MnPt | 139 | FM | Hg2YMn | 225 | AFM | Fe2NbRh | 119 | FM |
| Pt2CrCo | 139 | FM | Ru2ScCr | 225 | AFM | Fe2ZrAu | 119 | FM |
| Sc2MnZn | 139 | FM | Ru2TiCr | 225 | AFM | Fe2MnIr | 119 | AFM |
| Ta2FePt | 139 | FM | Ti2MoMn | 225 | AFM | Co2MnIr | 119 | AFM |

Table S. 6- List of all-*d*-metal Heusler that have a stable martensite and meta-stable austenite with their ground state.

| Comp. | Mom. Mart. [μB] | a Mart. [Å] | c/a Mart | Mom. Aust. [μB] | a Aust. [Å] | SPG |
|---|---|---|---|---|---|---|
| Co2TaV | 2.8,2.8,0.3,0.3,0.3,0.3,-0.3,-0.3 | 3.82 | 1.32 | 3.5,3.5,0.4,0.4,0.4,0.4,0,0 | 4.24 | 225 |
| Co2NbV | 0,0,0,0,0,0,0,0 | 3.81 | 1.32 | 1.2,1.2,1.3,1.3,1.3,1.3,0,0 | 4.24 | 225 |
| Co2VMn | 0,0,0,0,0,0,0,0 | 3.53 | 1.42 | 1.1,1.1,1.3,1.3,1.3,1.3,0,1.0,1 | 4.07 | 225 |
| Co2TiMo | -0.3,-0.3,0,0,0,0,1.2,1.2 | 3.82 | 1.32 | 0.1,0.1,1.3,1.3,1.3,1.3,3.1,3.1 | 4.24 | 225 |
| Co2VFe | 0,0,0,0,0,0,0,0 | 3.58 | 1.39 | 0.6,0.6,1.4,1.4,1.4,1.4,0.0,1.0,1 | 4.05 | 225 |
| Co2TaFe | 2.1,2.1,1.0,1.0,1.0,1.0,-0.8,-0.8 | 3.73 | 1.36 | 2.8,2.8,1.5,1.5,1.5,1.5,0.2,0.2 | 4.19 | 225 |
| Co2NbFe | 2.1,2.1,0.8,0.8,0.8,0.8,-0.3,-0.3 | 3.72 | 1.37 | 2.9,2.9,1.4,1.4,1.4,1.4,-0.1,-0.1 | 4.19 | 225 |
| Zn2TiMn | 2.2,2.2,0.8,0.8,0.8,0.8,-0.3,-0.4 | 3.75 | 1.52 | 2.9,2.9,1.5,1.5,1.5,1.5,0,0 | 4.35 | 225 |
| Zn2MnRh | 2.3,2.3,-0.1,-0.1,-0.1,-0.1,-0.9,-0.9 | 3.86 | 1.38 | 3.2,3.2,0,0,0,0,-0.4,-0.4 | 4.27 | 225 |
| Zn2MnPd | 3.7,3.7,0,0,0,0,0.4,0.4 | 4.00 | 1.27 | 3.4,3.4,0,0,0,0,0,0 | 4.32 | 225 |
| Zn2HfMn | 3.5,3.5,0,0,0,0,0,0 | 3.92 | 1.45 | 3.5,3.5,0,0,0,0,0,0 | 4.47 | 225 |
| Zn2CrPt | 2.2,2.2,-0.1,-0.1,-0.1,-0.1,-0.4,-0.4 | 3.98 | 1.31 | 3.2,3.2,0,0,0,0,-0.3,-0.3 | 4.33 | 225 |
| V2FeIr | 3.4,3.4,0,0,0,0,0.1,0.1 | 3.80 | 1.34 | 3.2,3.2,0,0,0,0,0,0 | 4.21 | 225 |
| V2FeCo | 1.5,1.5,-0.3,-0.3,-0.3,-0.3,0,0 | 3.82 | 1.20 | 1.9,1.9,0.2,0.2,0.2,0.2,0.1,0.1 | 4.08 | 225 |
| V2RuCo | -0.3,-0.3,0.2,0.2,0.2,0.2,0.6,0.6 | 3.94 | 1.19 | 1.0,1.0.2,0.2,0.2,0.2,1.2,1.2 | 4.18 | 225 |
| V2IrCo | 0.7,0.7,0.3,0.3,0.3,0.3,-0.1,-0.1 | 3.78 | 1.35 | 1.2,1.2,0.6,0.6,0.6,0.6,0,0 | 4.22 | 225 |
| Ta2FeRh | 0,0,0,0,0,0,0,0 | 4.12 | 1.23 | 1.4,1.4,0.9,0.9,0.9,0.9,0.1,0.1 | 4.44 | 225 |
| Ta2FeIr | 0.1,0.1,0,0,0,0,0,0 | 4.13 | 1.23 | 1.7,1.7,-0.1,-0.1,-0.1,-0.1,0.2,0.2 | 4.46 | 225 |
| Ta2CrPt | 0.5,0.5,0,0,0,0,0,0 | 3.98 | 1.46 | 1.9,1.9,0,0,0,0,0.2,0.2 | 4.49 | 225 |
| Pt2MnCu | 2.3,2.3,-0.1,-0.1,-0.1,-0.1,0,0 | 3.96 | 1.31 | 0,0,0,0,0,0,0,0 | 4.33 | 225 |
| Pd2MnAg | 3.6,3.6,0.1,0.1,0.1,0.1,0,0 | 4.14 | 1.26 | 3.7,3.7,0.2,0.2,0.2,0.2,0,0 | 4.48 | 225 |
| Ni2TaMn | 4.0,4.0.1,0.1,0.1,0.1,0,0 | 3.67 | 1.46 | 4.0,4.0.2,0.2,0.2,0.2,0,0 | 4.24 | 225 |
| Ni2NbMn | 2.2,2.2,0.3,0.3,0.3,0.3,-0.3,-0.3 | 3.67 | 1.47 | 3.5,3.5,0.3,0.3,0.3,0.3,0,0 | 4.24 | 225 |
| Ni2MnCu | 2.3,2.3,0.3,0.3,0.3,0.3,-0.4,-0.4 | 3.67 | 1.34 | 3.6,3.6,0.3,0.3,0.3,0.3,0,0 | 4.05 | 225 |
| Ni2VMn | 2.6,-2.6,0,0,0,0,1.0,-1.0 | 3.56 | 1.46 | 3.3,3.3,0.5,0.5,0.5,0.5,0.3,0.3 | 4.11 | 225 |
| Ir2MnCu | 3.2,3.2,0.4,0.4,0.4,0.4,0.4,0,0 | 3.85 | 1.32 | 3.3,3.3,0.5,0.5,0.5,0.5,0,0 | 4.22 | 225 |
| Fe2TiNi | -0.4,-0.4,2.3,2.3,2.3,2.3,2.3,0,0 | 3.55 | 1.58 | -0.5,-0.5,2.4,2.4,2.4,2.4,0,0 | 4.11 | 225 |
| Fe2CrRh | 0.4,0.4,2.3,2.3,2.3,2.3,-0.4,-0.4 | 3.69 | 1.42 | 0.7,0.7,2.0,2.0,2.0,2.0,-0.7,-0.7 | 4.14 | 225 |
| Fe2CrPt | -2.2,-2.2,2.4,2.4,2.4,2.4,0.3,0.3 | 3.61 | 1.44 | -0.6,-0.6,2.8,2.8,2.8,2.8,0.8,0.8 | 4.18 | 225 |
| Cu2ZrMn | -2.6,-2.6,2.5,2.5,2.5,2.5,0.1,0.1 | 3.81 | 1.49 | -0.4,-0.4,2.7,2.7,2.7,2.7,0.4,0.4 | 4.44 | 225 |
| Cu2TiMn | 2.7,2.7,0,0,0,0,-0.4,-0.4 | 3.63 | 1.55 | 3.8,3.8,0.1,0.1,0.1,0.1,0,0 | 4.28 | 225 |
| Cu2HfMn | 2.4,2.4,0,0,0,0,-0.6,-0.6 | 3.79 | 1.49 | 3.6,3.6,0.1,0.1,0.1,0.1,0,0 | 4.41 | 225 |
| Cu2ScFe | 2.5,2.5,0,0,0,0,-0.3,-0.3 | 3.77 | 1.49 | 3.7,3.7,0.1,0.1,0.1,0.1,0,0 | 4.35 | 225 |
| Au2MnCu | 2.1,2.1,0,0,0,0,-0.3,-0.3 | 4.18 | 1.22 | 2.6,2.6,0,0,0,0,-0.2,-0.2 | 4.46 | 225 |
| Zn2CrRh | 3.9,3.9,0,0,0,0,0,0 | 3.88 | 1.37 | 3.9,3.9,0,0,0,0,0,0 | 4.31 | 225 |
| Fe2IrCo | 3.3,3.3,0,0,0,0,0.2,0.2 | 3.70 | 1.35 | 3.3,-3.3,0,0,0,0,-0.1,0.1 | 4.16 | 225 |
| Fe2VIr | 0,0,2.5,-2.5,2.5,-2.5,0,0 | 3.70 | 1.40 | 1.9,1.9,3.0,3.0,3.0,3.0.9,0.9 | 4.18 | 225 |
| Ir2MnZn | 0,0,2.0,2.0,-2.0,-2.0,0,0 | 3.93 | 1.28 | -1.2,-1.2,2.4,2.4,2.4,2.4,0.6,0.6 | 4.27 | 225 |
| Mn2IrCo | 2.8,-2.8,0,0,0,0,0,0 | 3.74 | 1.33 | 3.0,3.0,0,0,0,0,0,0 | 4.18 | 225 |
| Mn2CoNi | 0,0,2.7,2.7,-2.7,-2.7,0,0 | 3.58 | 1.40 | 1.4,1.4,3.1,3.1,3.1,3.1,0.3,0.3 | 4.10 | 225 |
| Mn2RhCo | 0,0,2.9,2.9,-2.9,-2.9,0,0 | 3.69 | 1.39 | 1.5,1.5,3.3,3.3,3.3,3.3,0.9,0.9 | 4.17 | 225 |
| Mn2FeIr | 0,0,2.8,2.8,-2.8,-2.8,0,0 | 3.66 | 1.41 | 1.3,1.3,3.2,3.2,3.2,3.2,0.5,0.5 | 4.15 | 225 |
| Mn2FePt | 0,0,2.5,2.5,-2.5,-2.5,0,0 | 3.73 | 1.38 | 1.4,1.4,2.7,2.7,2.7,2.7,0.2,0.2 | 4.21 | 225 |
| Mn2NbIr | 0,0,2.9,2.9,-2.9,-2.9,0,0 | 3.82 | 1.45 | 1.3,1.3,3.1,3.1,3.1,3.1,0.3,0.3 | 4.32 | 225 |
| Mn2MoPt | 0,0,2.5,-2.5,2.5,-2.5,0,0 | 3.89 | 1.34 | 0.3,0.3,2.6,2.6,2.6,2.6,-0.7,-0.7 | 4.33 | 225 |
| Mn2CoPt | 0,0,2.7,2.7,-2.7,-2.7,0,0 | 3.75 | 1.38 | -0.9,-0.9,3.2,3.2,3.2,3.2,0.3,0.3 | 4.25 | 225 |
| Mn2RhPt | 0,0,3.1,3.1,-3.1,-3.1,0,0 | 3.90 | 1.33 | 0.4,0.4,3.5,3.5,3.5,3.5,1.6,1.6 | 4.36 | 225 |
| Mn2ReIr | 0,0,3.4,3.4,-3.4,-3.4,0,0 | 3.80 | 1.37 | 0.4,0.4,3.8,3.8,3.8,3.8,0.8,0.8 | 4.26 | 225 |
| Mn2RhNi | 0,0,2.4,2.4,-2.4,-2.4,0,0 | 3.72 | 1.39 | -0.6,-0.6,3.0,3.0,3.0,3.0,0.4,0.4 | 4.24 | 225 |
| Rh2TiCr | 0,0,3.2,3.2,-3.2,-3.2,0,0 | 4.04 | 1.23 | 0.9,0.9,3.6,3.6,3.6,3.6,0.7,0.7 | 4.35 | 225 |

| Composition | | | | | | |
|---|---|---|---|---|---|---|
| Rh2MnFe | 2.7,-2.7,0,0,0,0,-0.1,0.1 | 3.84 | 1.33 | 3.0,3.0.4.0.4,0.4,0.4,0,0 | 4.29 | 225 |
| Rh2MnCu | 3.1,-3.1,0,0,0,0,2.9,-2.9 | 3.85 | 1.32 | 3.7,3.7,0.9,0.9,0.9,0.9,3.3,3.3 | 4.23 | 225 |
| Fe2CrIr | 3.0,-3.0.0.0.0.0,-0.1,0.1 | 3.67 | 1.43 | 3.1,-3.1,0,0,0,0,-0.1,0.1 | 4.15 | 219 |
| Fe2HfPt | -2.0,-2.0,2.4,2.3,2.4,2.3,0.2,0.2 | 3.80 | 1.48 | 1.0,1.0,2.8,2.4,2.8,2.4,0.4,0.4 | 4.41 | 219 |
| Fe2MoIr | -0.3,-0.3,2.3,2.4,2.3,2.4,0.1,0.1 | 3.73 | 1.44 | -0.2,-0.2,2.9,2.2,2.9,2.2,0.2,0.2 | 4.26 | 219 |
| Fe2ReIr | -0.5,-0.5,2.2,1.9,2.2,1.9,0.1,0.1 | 3.71 | 1.43 | -0.2,-0.2,2.8,2.3,2.8,2.3,0.4,0.4 | 4.25 | 219 |
| Fe2NbRh | -0.4,-0.4,2.1,1.7,2.1,1.7,-0.1,-0.1 | 3.76 | 1.47 | 0,0,2.7,2.6,2.7,2.6,0.5,0.5 | 4.29 | 219 |
| Fe2VRh | 0.2,0.2,2.3,2.2,2.3,2.2,-0.4,-0.4 | 3.68 | 1.44 | 0.4,0.4,1.8,2.8,1.8,2.8,-0.2,-0.2 | 4.16 | 219 |
| Fe2TaIr | 0.3,0.3,2.3,2.3,2.3,2.3,-1.0,-1.0 | 3.78 | 1.46 | 0.4,0.4,1.9,2.8,1.9,2.8,-0.6,-0.6 | 4.32 | 219 |
| Fe2TiPt | -0.3,-0.3,2.1,2.1,2.1,2.1,0.2,0.2 | 3.75 | 1.47 | -0.2,-0.2,2.9,2.0,2.9,2.0,0.4,0.4 | 4.28 | 219 |
| Fe2PtZn | -0.6,-0.6,2.4,2.4,2.4,2.4,0.1,0.1 | 3.85 | 1.33 | -0.3,-0.3,2.9,2.1,2.9,2.1,0.2,0.2 | 4.23 | 219 |
| Mn2WIr | -0.6,-0.6,2.5,2.9,2.5,2.9,-0.1,-0.1 | 3.73 | 1.48 | 0.1,0.1,2.4,-0.6,2.4,-0.6,0,0 | 4.29 | 219 |
| Ti2MnAu | 0,0,-1.9,2.3,-1.9,2.3,0.1,0.1 | 3.91 | 1.50 | -0.2,-0.2,3.3,1.2,3.3,1.2,0.3,0.3 | 4.47 | 219 |
| Fe2MnIr | 2.8,2.8,-0.7,-0.7,-0.7,-0.7,0,0 | 3.68 | 1.37 | 3.3,3.3,-0.5,-1.0,-0.5,-1.0,1.0,0.1 | 4.16 | 219 |
| Fe2MnPt | 2.5,-2.5,2.3,0,-2.3,0,0,0 | 3.72 | 1.37 | 2.6,2.6,2.7,2.2,2.7,2.2,0.5,0.5 | 4.22 | 219 |
| Mn2TaPt | 2.8,-2.8,2.5,0,-2.5,0,0,0 | 3.78 | 1.51 | 3.3,3.3,2.9,2.4,2.9,2.4,0.4,0.4 | 4.36 | 219 |
| Co2MnIr | 0,0,2.8,-2.1,-2.8,2.1,-0.3,0.3 | 3.61 | 1.42 | 0.2,0.2,1.3,3.5,1.3,3.5,-0.1,-0.1 | 4.16 | 219 |
| Mn2TiPt | 2.4,-2.4,-0.4,0,0.4,0,0,0 | 3.79 | 1.46 | 3.2,3.2,1.9,1.8,1.9,1.8,0.7,0.7 | 4.35 | 219 |

Table S. 7- Data for plot of Figure 4 (a).

| Composition | $T_{C/N}^{mart}$ [k] | $T_{C/N}^{aust}$ [k] | $T_M$ [k] | $T_M$ with $F^{mag}$ [k] |
|---|---|---|---|---|
| Au2MnCu | -260 | -220 | 50 | 50 |
| Zn2MnPd | -415 | 585 | 370 | 360 |
| Rh2MnCu | -370 | -20 | 545 | 505 |
| Cu2ZrMn | 735 | 275 | 585 | 550 |
| Zn2HfMn | 315 | 0 | 700 | 830 |
| Cu2ScFe | 545 | 285 | 750 | 740 |
| Mn2TiPt | -465 | 375 | 860 | 850 |
| Pd2MnAg | -175 | -35 | 890 | --- |
| Zn2CrRh | 100 | -90 | 980 | 1160 |
| Ni2NbMn | 665 | 325 | 1220 | 1205 |
| Fe2VIr | -510 | 295 | 1260 | 1220 |
| Mn2NbIr | -630 | 995 | --- | --- |
| Cu2TiMn | 875 | 385 | --- | --- |
| Fe2CrIr | -660 | 685 | --- | --- |
| Fe2HfPt | 0 | 20 | --- | --- |
| Fe2CoIr | -575 | 955 | --- | --- |
| Fe2MoIr | 270 | 485 | --- | --- |
| Fe2ReIr | 315 | 615 | --- | --- |
| Fe2VRh | 115 | 835 | --- | --- |
| Fe2TaIr | 415 | 935 | --- | --- |
| Fe2TiPt | 800 | 945 | --- | --- |
| Fe2PtZn | 600 | 745 | --- | --- |
| Fe2MnPt | -310 | 600 | --- | --- |
| Ir2MnCu | -70 | -35 | --- | --- |
| Ir2MnZn | -170 | 35 | --- | --- |
| Mn2CoIr | -1275 | 635 | --- | --- |
| Mn2CoNi | -1290 | 885 | --- | --- |
| Mn2CoRh | -1320 | 700 | --- | --- |
| Mn2FeIr | 1225 | 215 | --- | --- |
| Mn2MoPt | -1040 | 295 | --- | --- |
| Mn2CoPt | -1250 | 815 | --- | --- |
| Mn2RhPt | -1120 | 215 | --- | --- |
| Mn2TaPt | -1050 | 200 | --- | --- |
| Mn2ReIr | -1110 | 195 | --- | --- |
| Mn2RhNi | -1250 | 395 | --- | --- |
| Ni2MnCu | 115 | 435 | --- | --- |
| Ni2TaMn | 70 | 300 | --- | --- |
| Pt2MnCu | 0 | 50 | --- | --- |
| Rh2MnFe | -575 | 745 | --- | --- |

| | | | |
|---|---|---|---|
| Ti2MnAu | 655 | 1015 | --- | --- |
| Zn2CrPt | 715 | -930 | --- | --- |
| Zn2MnRh | 400 | 325 | --- | --- |
| Zn2TiMn | 865 | 625 | --- | --- |
| Co2NbFe | 415 | 535 | --- | --- |
| Co2TaFe | 435 | 420 | --- | --- |
| Co2VFe | 485 | 755 | --- | --- |
| Co2VMn | -240 | 885 | --- | --- |
| Mn2WIr | -445 | 325 | --- | --- |
| Ni2VMn | 30 | 340 | --- | --- |

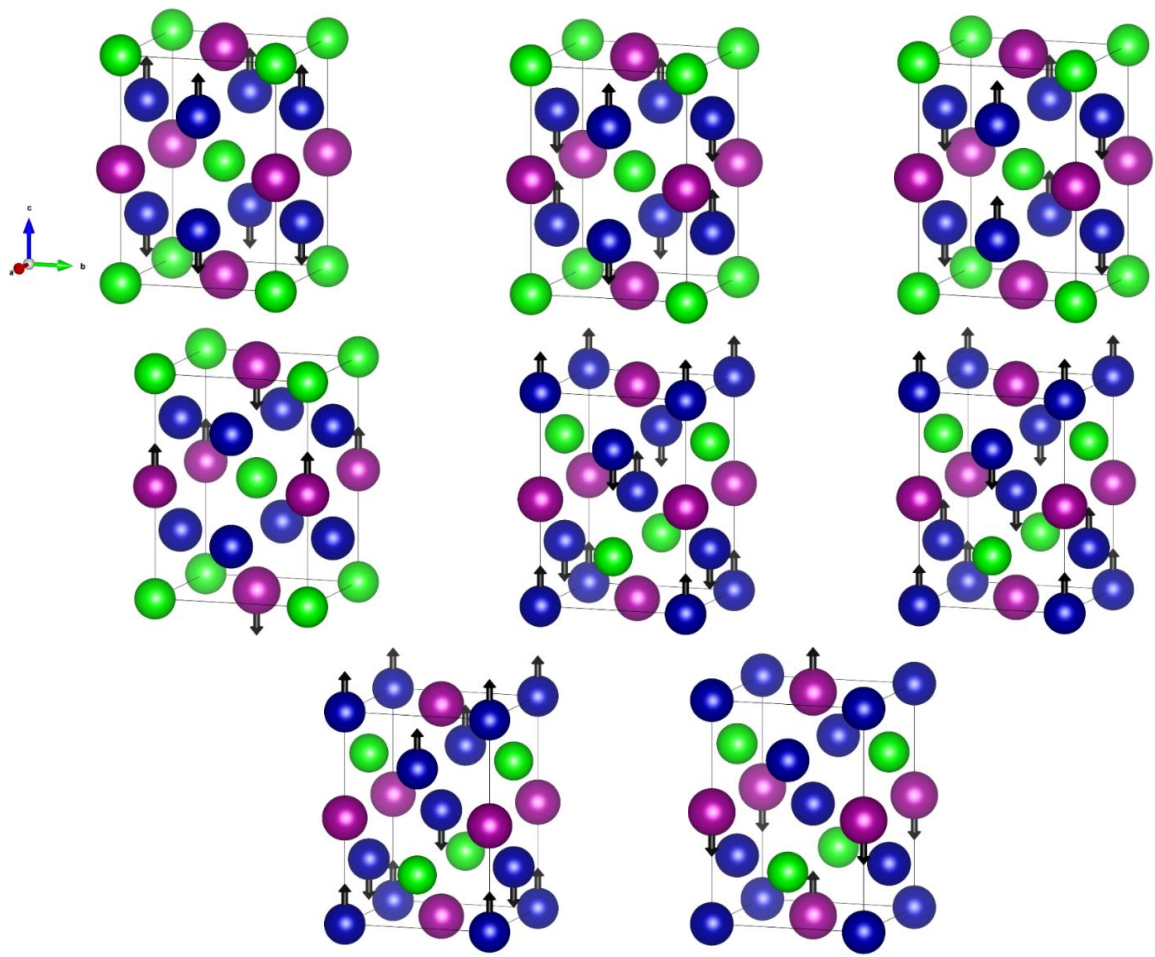

Figure S. 1- Spin configurations considered in the AFM screen procedure, for both the regular and the inverse phases.

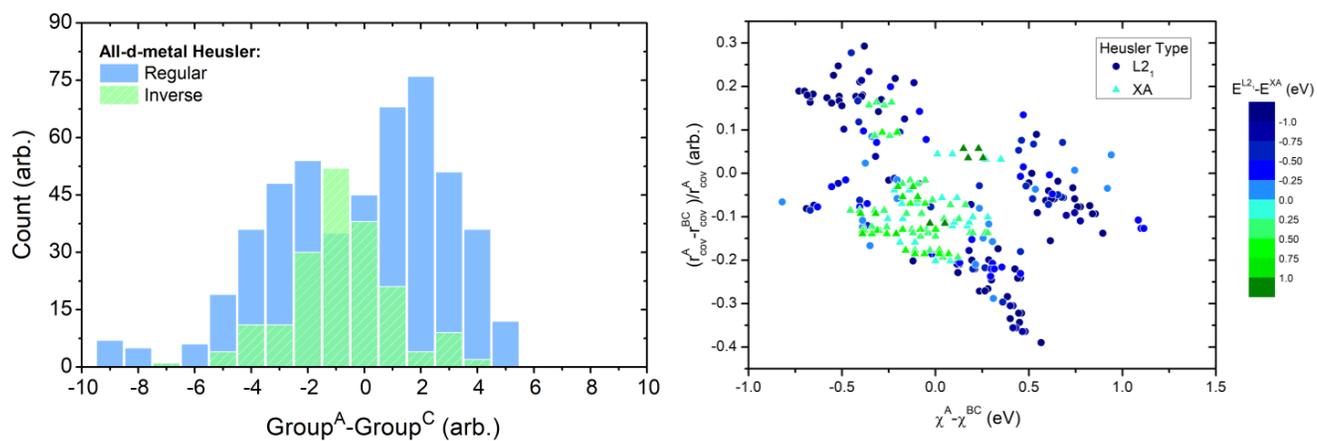

Figure S. 2- Left: The distribution of regular and inverse structures for the all-*d*-metal Heuslers as function of the difference of group number of site A and C. Right: The radius mismatch and electronegativity map for cubic Heusler structures.

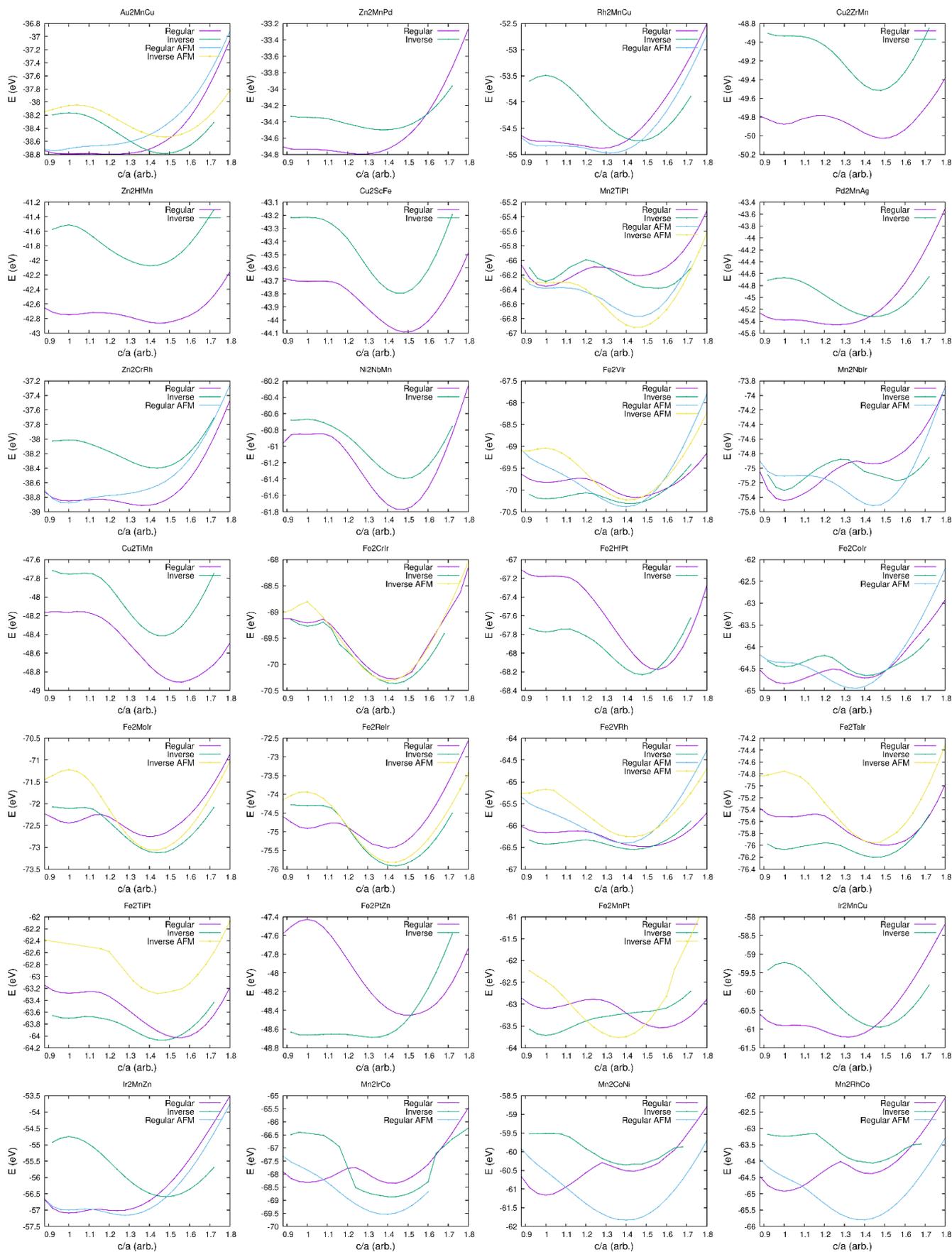

Figure S.3 – Bain Paths of Heuslers with possible structural transition, with AFM states when applicable (Part 1).

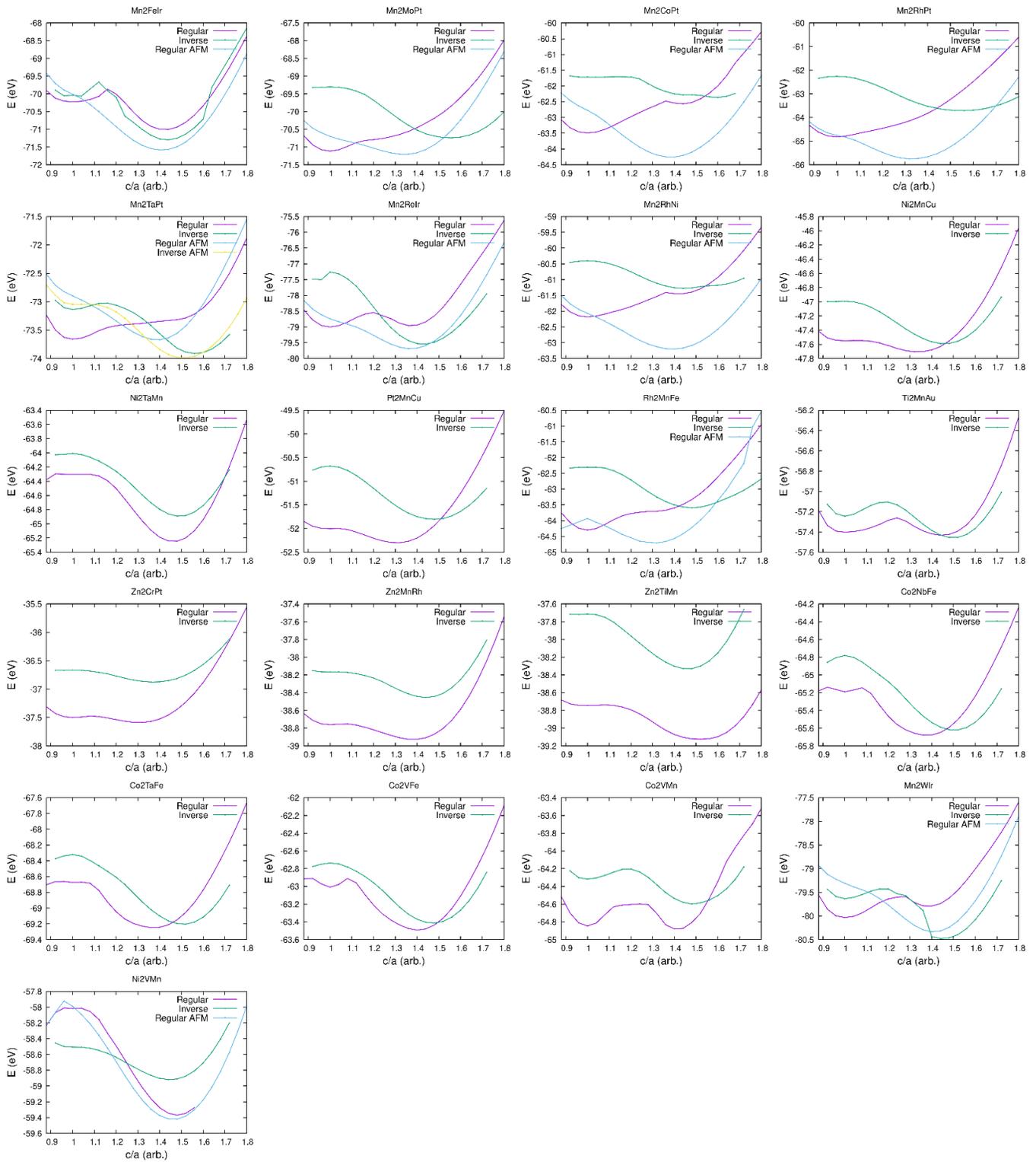

Figure S.4 – Bain Paths of Heuslers with possible structural transition, with AFM states when applicable (Part 2).


\end{document}